\documentclass[journal]{IEEEtran}
\usepackage[utf8]{inputenc}
\usepackage[T1]{fontenc}
\usepackage{algorithm}
\usepackage{algorithmicx} 
\usepackage[noend]{algcompatible}
\usepackage{amsmath}
\usepackage{amsthm}
\usepackage{amssymb}
\usepackage{graphicx}
\usepackage{setspace}
\usepackage{epstopdf}
\usepackage[font={normal}]{caption}
\usepackage{cite}
\usepackage{bm}
\usepackage{bbm}
\usepackage{algpseudocode}
\usepackage{array,multirow,makecell}\setcellgapes{1pt}
\usepackage{comment}
\usepackage{enumitem}
\usepackage{subfigure}
\usepackage[table]{xcolor}
\usepackage{balance}
\usepackage{mathtools}

\DeclarePairedDelimiter\floor{\lfloor}{\rfloor}
\DeclareMathOperator*{\argmax}{argmax}

\newcolumntype{R}[1]{>{\raggedleft\arraybackslash }b{#1}}
\newcolumntype{L}[1]{>{\raggedright\arraybackslash }b{#1}}
\newcolumntype{C}[1]{>{\centering\arraybackslash }b{#1}}
\makegapedcells
\allowdisplaybreaks
\makeatletter
\def\maketag@@@#1{\hbox{\m@th\normalfont\normalsize#1}}
\makeatother
\begin{document}

\title{LoRa-RL: Deep Reinforcement Learning for Resource Management in Hybrid Energy LoRa Wireless Networks}
\author{Rami~Hamdi,~\IEEEmembership{Member,~IEEE,}
	Emna~Baccour,
	Aiman~Erbad,~\IEEEmembership{Senior~Member,~IEEE,}
	Marwa~Qaraqe,~\IEEEmembership{Member,~IEEE,}
	and~Mounir~Hamdi,~\IEEEmembership{Fellow,~IEEE}
\thanks{The authors are with the Division of Information and Computing Technology, College of Science and Engineering, Hamad Bin Khalifa University, Qatar Foundation, Doha, Qatar (email: hrami@hbku.edu.qa; ebaccourepbesaid@hbku.edu.qa; aerbad@ieee.org; mqaraqe@hbku.edu.qa; mhamdi@hbku.edu.qa ).}
}
\maketitle
\thispagestyle{empty}

\begin{abstract}
LoRa wireless networks are considered as a key enabling technology for next generation internet of things (IoT) systems. New IoT deployments (e.g., smart city scenarios) can have thousands of devices per square kilometer leading to huge amount of power consumption to provide connectivity. In this paper, we investigate green LoRa wireless networks powered by a hybrid of the grid and renewable energy sources, which can benefit from harvested energy while dealing with the intermittent supply. This paper proposes resource management schemes of the limited number of channels and spreading factors (SFs) with the objective of improving the LoRa gateway energy efficiency. First, the problem of grid power consumption minimization while satisfying the system's quality of service demands is formulated. Specifically, both scenarios the uncorrelated and time-correlated channels are investigated. The optimal resource management problem is solved by decoupling the formulated problem into two sub-problems: channel and SF assignment problem and energy management problem. Since the optimal solution is obtained with high complexity, online resource management heuristic algorithms that minimize the grid energy consumption are proposed. Finally, taking into account the channel and energy correlation, adaptable resource management schemes based on Reinforcement Learning (RL), are developed. Simulations results show that the proposed resource management schemes offer efficient use of renewable energy in LoRa wireless networks.
\end{abstract}
\begin{IEEEkeywords}
LoRa, energy harvesting, resource management, reinforcement learning.
\end{IEEEkeywords}

\section{Introduction}
LoRa wireless networks are considered as a key technology for next generation of internet of things (IoT) wireless networks~\cite{survey}. These systems are based on the deployment of a large number of low-powered connected devices. Indeed, innovative wireless network, such as LoRa, enables the exponential growth connected devices, robust operations, wider coverage, and higher energy efficiency~\cite{survey}. Hence, LoRa may provide sustainable connectivity to low-powered devices distributed over very large geographical areas~\cite{survey2,ref1}. LoRa which operates in the unlicensed bands~\cite{ref2} provides also adaptive transmission rates and coverage for low-powered devices. LoRa enables long range transfer of information with a low transfer rate.~\cite{ref4}. The chirp spreading modulation (CSM) was adopted as the modulation technique for LoRa transmission~\cite{ref5}. This scheme is based on coding the information in the frequency shift at the beginning of the symbol. The chirp is assumed to be as a kind of carrier and the modulated signal is a chirp waveform which its behaviour depends on the SF. LoRa signals with different SFs are quasi-orthogonal~\cite{ref5}. However, LoRa signals with the same SF exhibit cross-correlation properties that could make them vulnerable to interference. The performance of CSM was theoretically investigated in~\cite{ref5}. The performance analysis of this modulation scheme was extended  by considering various fading channels in~\cite{ref6} and by considering interference in~\cite{ref7}. The scalability of LoRa networks was investigated in~\cite{ref8} by proposing a stochastic geometry framework. This framework supports the exponential growth of connected devices. Furthermore, adequate and intelligent resource management strategies may be adopted in LoRa networks to enhance the system performance.

RL approaches have become increasingly popular, particularly for systems with complex and dynamic problem spaces~\cite{QLearning,DQN,reinforce}. These approaches are apt to act under unforeseen environments by making decisions, receiving rewards and penalties, and learning policies based on the system conditions. Unlike supervised learning, RL pursues the optimal solution by interacting with the environment parameters (e.g., the total required energy, and the current energy price). In particular, the RL approach adopts a trial-and-error search method to discover the network environment and learn the resource management policy without labeling the data at each time step. Learning the statistical distributions of the environment parameters produces the most effective action policy that adapts to changes over time and leads to the maximum reward. Hence, RL is powerful tool that can be applied in the resource management problem in LoRa wireless networks.

For greening and improving energy efficiency of LoRa networks, the devices may be powered by renewable energy sources~\cite{energy,energy2}. Energy harvesting allows wireless systems to continually acquire energy from nature or man-made phenomena (solar, wind, electromagnetic, ...). It provides wireless devices self-sustainability and virtually perpetual operation. Indeed, it allows reducing the use of conventional energy and accompanying carbon footprint. Hence, we propose to investigate in this work a resource management problem in hybrid energy LoRa wireless networks. We propose various resource management schemes considering both scenarios uncorrelated and time-correlated channels. An optimal resource management solution with high complexity is proposed as benchmark. Low complexity heuristic resource management schemes are proposed for the case of real-time application. Moreover, smart and adaptable resource management approaches based on RL were developed. Hence, the key contributions are summarized as follows:
\begin{itemize}
\item We formulated the problem of grid energy cost minimization in LoRa wireless networks while the LoRa gateway (LG) is powered by both an energy harvesting source and the grid.
\item We solved the optimal offline resource management problem by decoupling the formulated problem into two sub-problems. The first one is a channel and SF assignment problem and the second one is an energy management problem.
\item We developed an optimal SF assignment scheme that minimizes the grid energy cost.
\item We investigated the online resource management problem by proposing  efficient heuristic channel, SF assignment, and energy management algorithms for both scenarios uncorrelated and time-correlated channels.
\item We proposed an efficient heuristic channel, SF assignment, and energy management algorithm for hybrid energy powered LoRa networks considering \textit{time-correlated} channels.
\item We developed our resource management algorithm using deep  reinforcement learning to reduce the complexity of the NP-hard optimization and implement an adaptive online energy assignment.
\item We performed extensive evaluation of the proposed resource management schemes under different scenarios to illustrate the system performance in terms of grid energy cost.
\end{itemize}

The remainder of the paper is organized as follows: The system model is presented in Section II. The related works are discussed in Section III. The resource management problem in hybrid energy LoRa networks is formulated in Section IV. The optimal offline resource management problem is investigated in Section V. The online resource management algorithms are developed in Section VI. The evaluation results are presented and discussed in Section VII. Finally, conclusions are provided in Section VIII.
\begin{figure}[!h]
	\centerline{\includegraphics [width=1\columnwidth]  {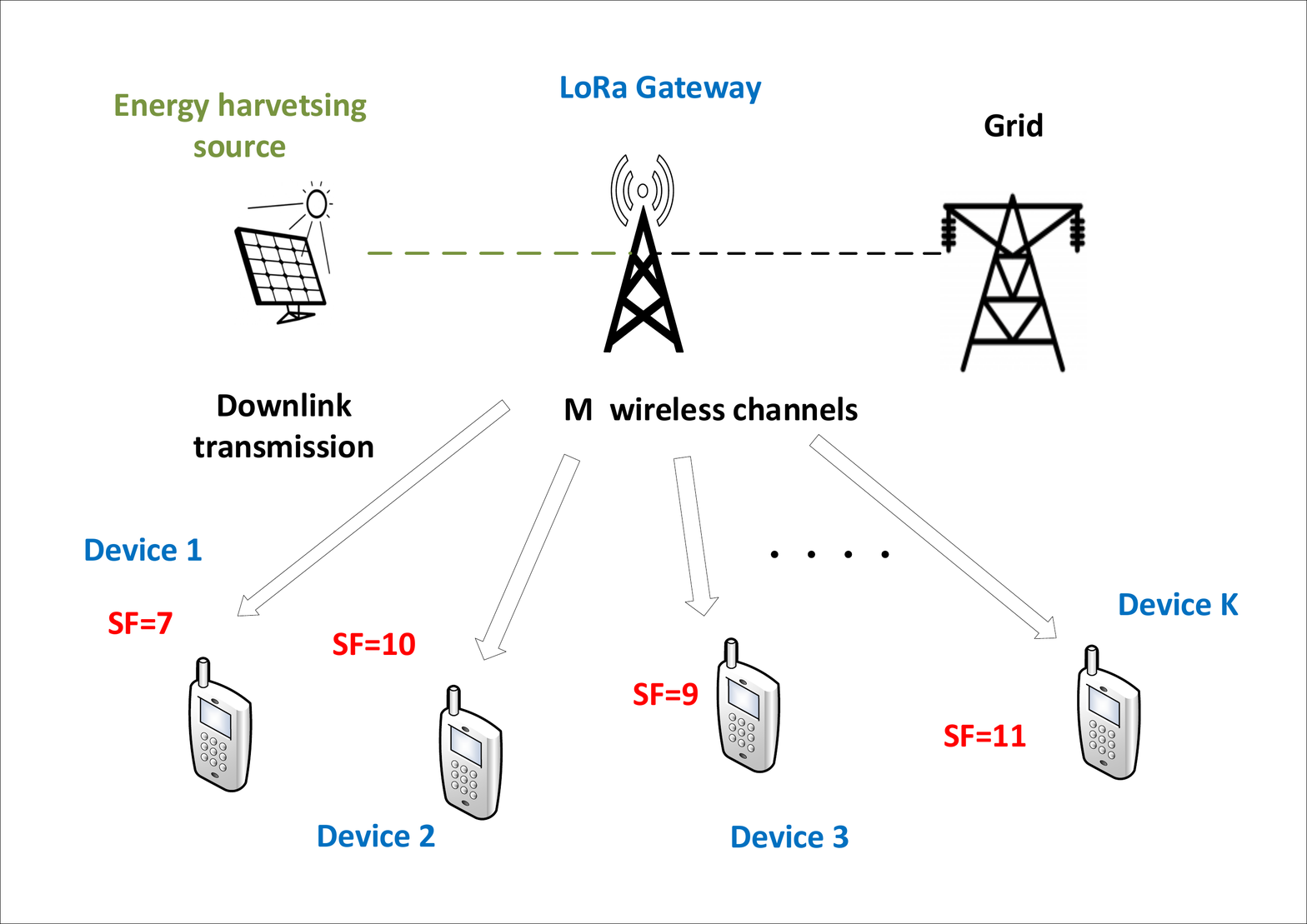}}
	\caption{Hybrid energy powered LoRa wireless networks.}
	\label{figs}
\end{figure}
\section{Related Work} 
Various works tackled the resource management problem under different LoRa network architectures and assumptions. Specifically, SF assignment, sub-band selection, user scheduling, and power allocation in LoRa networks were the focus of~\cite{icc20,ress1,ress2,ress3,ress4,ress5,ress6,ress7,ress8,ress9,ress10,ress11,ress12,ress13,icc21,ress14,ress15,ress16,ress17,ress18,ress19,new1,new2,new3,new4,new5}. An efficient SF assignment scheme was developed in~\cite{icc20} based on instantaneous channel realizations to enhance the symbol error rate. Also, the authors of~\cite{ress1} proposed a novel SF allocation strategy based on matching theory to optimize the LoRa network throughput. Moreover, the number of connected devices was optimized in~\cite{ress2} by proposing an efficient SF assignment scheme. A power allocation and SF assignment solution was proposed in~\cite{ress3} to reduce energy consumption in LoRa networks with imperfect SF orthogonality. An energy-efficient user scheduling, SF assignment, and power allocation scheme was proposed in~\cite{ress4}. In~\cite{ress5,ress6}, the authors proposed efficient SF assignment schemes to maximize the packet success probability. The authors of~\cite{ress7} proposed an efficient interference-aware SF assignment. An adequate SF allocation strategy was proposed in~\cite{ress8} to reduce the convergence period in LoRa networks with an adaptive data rate mechanism. In~\cite{ress9}, a multi hop LoRa system was investigated in order to enable energy-efficient connectivity in smart city applications and the system performance was evaluated based on a experimental case study. This system was investigated further in~\cite{ress10} by proposing an efficient clustering algorithm. A capacity maximization problem in LoRa networks was studied in~\cite{ress11}. User scheduling was incorporated to a multi channel LoRa network in~\cite{ress12} to improve the synchronization packet length. The average number of decoded LoRa frames was investigated in~\cite{ress13} by taking into account physical layer and medium access control. In~\cite{icc21}, the devices are powered by energy harvesting sources for uplink transmission considering only one channel. The optimal energy management and SF assignment algorithm was devised. In~\cite{ress14}, the authors proposed a low complexity energy-efficient rate control scheme for LoRa uplink transmission based on Markov chain. In~\cite{ress15}, the authors investigated the performance of LoRa networks in term of latency by proposing a sub-band selection scheme. The authors of~\cite{ress16} provided a theoretical analysis of the achievable LoRa throughput in uplink considering imperfect SF orthogonality. In~\cite{ress17}, the trade-off relation between the waiting time and the energy consumption in LoRa networks was optimized by deriving the optimal number of ping slots. In~\cite{ress18}, the authors proposed an appropriate radio configuration scheme based on integer linear programming, which takes into account the scalability of the network in order to enhance the reliability of the LDs. A dynamic LoRa transmission control system was proposed in~\cite{ress19} to improve the energy efficiency. In~\cite{new1}, the LoRa capacity was enhanced by proposing an interleaved chirp spreading scheme. Moreover, the LoRa goodput was improved in~\cite{new2} by controlling the receiver window size. In~\cite{new3}, the coverage in hybrid LoRa networks was enhanced by developing an optimal planning scheme. The energy efficiency of LoRa networks was improved in~\cite{new4} by proposing a tree network adopted for LoRa to mitigate the energy consumption constraints. In~\cite{new5}, the authors proposed a novel fair and scalable relay control scheme for LoRa wireless networks to improve the success probability.\\
Different from the existing works, we proposed in this paper to investigate a novel system model for LoRa wireless networks powered by both an energy harvesting source and the grid, which can benefit from harvested energy while dealing with the intermittent supply. The design of energy-efficient LoRa networks is challenging due to the intermittency of renewable energy sources. Since most of the existing works are dealing with the coverage and the throughput, we formulated in this paper a novel challenging optimization problem with the objective of improving the LoRa gateway energy efficiency. Furthermore, we developed smart and adaptive resource management schemes for green wireless networks based on deep reinforcement learning.

\begin{table*}[t]
	\centering
	\caption{Summary of Important Notations.}
	\label{tab0}
	\begin{tabular}{|C{2cm}|L{4.4cm}||C{2cm}|L{5.7cm}|}
		\hline   \textbf{Symbol} & \textbf{Description} & \textbf{Symbol} & \textbf{Description}\\
		\hline $K$ & Number of LDs & $p_{k}(i)$ & Power allocated for LD $k$ \\
      \hline $B_m$ & Bandwidth of channel $m$ & $\sigma_m^2$ & Noise variance \\
       \hline $L$ & Number of frames & $\gamma_{k,m}(i)$ & SNR of LD $k$ through channel $m$ at frame $i$ \\
       \hline  $T_{out}$ & Duration of a frame & $B_{max}$ & Capacity of the battery \\
       \hline  $g_{k,m}(i)$ & Channel coefficient of LD $k$ through channel $m$ & $E(i)$ & Amount of harvested energy at frame $i$ \\
       \hline $\beta_k(i)$ & Path loss & $B(i)$ & Battery level at frame $i$ \\
      \hline $h_{k,m}(i)$ & Small-scale fading channel coefficient & $E_c$ & Fixed energy consumed by the circuit \\
     \hline $\alpha_k(i)$ & Spreading factor & $W_{i}$ & Grid's weight \\
     \hline $\Gamma$ & Set of SFs & $X^g(i)$ & Energy drawn from the power grid \\
      \hline $T$ & Duration of a sample transmission & $\gamma_{th}$ & Minimum received SNR \\
      \hline $\chi_{k,m}(i)$ & Channel assignment index & $\psi_m(i)$ & Set of devices assigned to channel $m$\\
	 \hline  $S$ & Set of states & $A$ & Set of possible actions \\
	 \hline $R$ & Immediate reward & $P$ & State transition probability \\
	 \hline $\gamma$ & Discount factor & $\theta$ & Neural network weights \\
	 \hline $\Pi$ & RL policy & $D$ & Experience memory \\
	 \hline $\lambda$ & Adjusting factor of PPO & $\epsilon$ & Cliprange of PPO \\
	 \hline $\rho$ & Soft update of DDPG & $N^s$ & Noise process of DDPG \\
	 \hline $\alpha $& Learning rate of the DNN &\multicolumn{2}{c|}{} \\
	 \hline
	\end{tabular}
\end{table*}

\section{System Model}
\subsection{Channel and Signal Model}
In this work, we consider a typical downlink LoRa wireless network (as shown in Fig.~\ref{figs}) that includes a LG serving K arbitrarily distributed LoRa devices (LDs) through M channels. Let $B_m$ denote the bandwidth of channel $m$. A given time interval is partitioned into $L$ frames with duration $T_{out}$. The channel coefficients between the gateway and LD $k$ through channel $m$ at frame $i$ is given by $g_{k,m}(i)=\beta_k(i) h_{k,m}(i)$, where $\beta_k(i)$ represents the path loss and $h_{k,m}(i)$  represents the small-scale fading channel coefficient. The important notations are summarized in Table~\ref{tab0}.\\\\ 
The LoRa modulation known as CSM is introduced in~\cite{ref5}, where the modulated signal is a chirp waveform and the frequency increases linearly with the time index. The LG sends a symbol $s_k(i)$ to LD $k$ at frame $i$ with duration $2^{\alpha_k(i)} T$, where $\alpha_k(i)$ is the SF taking values in $\Gamma=\{7,8,9,10,11,12\}$ and $T=\frac{T_{out}}{2^{12}}$ is the duration of a sample transmission~\cite{ref4}. The LG sends $\alpha_k(i)$ bits to device $k$ at frame $i$. The symbol $s_k(i)$ takes values in $\left\{0,1,2, \ldots, 2^{\alpha_k(i)}-1\right\}$~\cite{ref4}. The LDs adopt different SF for transmission in order to ensure orthogonality and enable multi user transmission~\cite{ref5}. Hence, the transmitted waveform vector for device $k$ at frame $i$ is given by:
\begin{equation}
\small
\label{eq:1}
\mathbf{x}_k(i)=  
\begin{bmatrix}
\left[ \frac{1}{\sqrt{2^{\alpha_k(i)}}} e^{\jmath 2 \pi\left[\left(s_k(i)+f\right)_{\bmod 2^{\alpha_k(i)}} \right] \frac{f}{2^{\alpha_k(i)}}}   \right]_{f=0..2^{\alpha_k(i)}-1} \\
\mathbf{0}_{2^{12}-2^{\alpha_k(i)}}
\end{bmatrix}
\end{equation} 
A zero padding with length $2^{12}-2^{\alpha_k(i)}$ is added for each vector in order to ensure the same vector length for all devices. The possible waveforms of CSM modulation are shown to be orthogonal~\cite{ref6}. Hence, the inner product receiver may be applied~\cite{ref5}. It consists of projecting the received vector $\mathbf{y}(i)$ onto the different signals given by:
\begin{equation}
\tiny
\label{eq:eq222}
\mathbf{c}_{|s_k(i)}=\left[ g^*_{k,m}(i)   \frac{1}{\sqrt{2^{\alpha_k(i)}}} e^{\jmath 2 \pi\left[\left(s_k(i)+f\right)_{\bmod 2^{\alpha_k(i)}} \right] \frac{f}{2^{\alpha_k(i)}}}   \right]^T_{f=0..2^{\alpha_k(i)}-1}
\end{equation}
and choosing the one with maximal square modulus projection. Hence, the best estimate of the transmitted signal $\hat{s}_k(i)$ by device $k$ at frame $i$ is given by:
\begin{equation}
\label{eq:2}
\hat{s}_k(i)= \argmax_{0..2^{\alpha_k(i)}-1} | \langle  \mathbf{y}(i),\mathbf{c}_{|s_k(i)} \rangle|^2.
\end{equation}
It is worth mentioning that only 6 LDs can be served simultaneously in one channel, since there are six available SFs range from 7 to 12~\cite{ref4}. Also, each LD can access at most one channel. Let $\chi_{k,m}(i)$ be a Boolean parameter that is set to 1 if LD $k$ at frame $i$ is assigned to channel $m$ and to 0 otherwise. Let $\psi_m(i)$ denote the set of devices assigned to channel $m$ at frame $i$. The vector of received signals through channel $m$ at frame $i$ is expressed as:
\begin{equation}
\label{eq:3}
\mathbf{y}_m(i)=  \sum_{k=1}^{K} \chi_{k,m}(i) p_{k}(i) g_{k,m}(i)  \mathbf{x}_k(i) + \mathbf{w}_m(i),
\end{equation}
where $p_{k}(i)$ is the power allocated for LD $k$ at frame $i$ and $\mathbf{w}_m(i)$ is assumed to be additive white Gaussian noise (AWGN) with zero mean and variance $\sigma_m^2$. Hence, the downlink signal-to-noise ratio (SNR) of LD $k$ through channel $m$ at frame $i$ is expressed as:
\begin{equation}
\label{eq:4}
\gamma_{k,m}(i)=\frac{ \chi_{k,m}(i) p_{k}(i) \mid \mathbf{g}_{k,m}(i) \mid^2}{\sigma_m^2}.
\end{equation}
The rate for LD $k$ through channel $m$ at frame $i$ is given by:
\begin{equation}
\label{eq:5}
R_{k,m}(i)= B_m \log_2 \left(1+\frac{ \chi_{k,m}(i) p_{k}(i) \mid \mathbf{g}_{k,m}(i) \mid^2}{\sigma_m^2} \right).
\end{equation}
An LD $k$ is scheduled at frame $i$, if it is assigned to one of the channels $\sum_{m=1}^{M}\chi_{k,m}(i)=1$, otherwise it is not scheduled and $\sum_{m=1}^{M}\chi_{k,m}(i)=0$.

\subsection{Energy Model}
\begin{figure}[!h]
	\centerline{\includegraphics [width=1\columnwidth]  {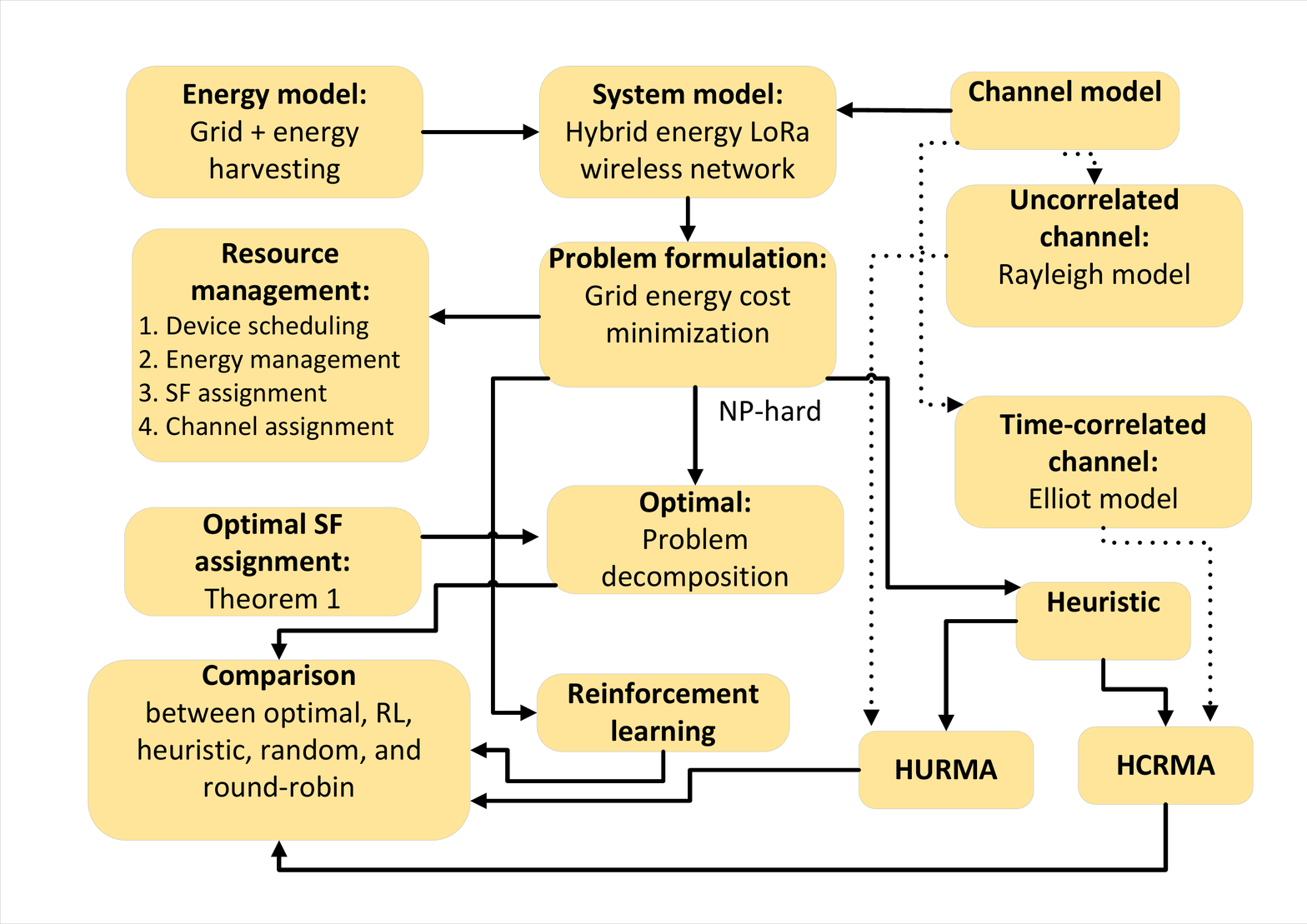}}
	\caption{Illustration of research design and different steps of the work.}
	\label{fig_new1}
\end{figure}

The LG is powered by both an energy harvesting source and the grid. The grid energy source compensates for the randomness and intermittency of the harvested energy. The harvested energy $E(i)$ is first stored in a battery with maximal capacity $B_{max}$. It is modeled as a correlated time process following a discrete-time Markov model as in~\cite{markov,markov2}, $E(i) \in \Omega \triangleq \{\omega_1,\omega_2, ..., \omega_M \}$ where $\Omega$ is the set of possible amount of harvested energy and $Q(\omega_m,\omega_j)=Pr(E(i+1)= \omega_m \mid E(i)= \omega_j )$ is the state transition probability. Let $B(i)$ denote the battery level at frame $i$. The required energy consumed at frame $i$ is given by:
\begin{equation}
\label{eq:6}
\begin{aligned}
X(i)&=X^h(i)+X^g(i)\\
& =E_c+  \sum_{k=1}^{K} p_{k}(i) 2^{\alpha_k(i)} T,
\end{aligned}
\end{equation}
where $E_c$ is a fixed energy consumed by the circuit which includes the amounts of power consumed by digital to analog converters (DACs), mixers and filters, and $X^h(i)$ and $X^g(i)$ denote the energy drawn from the energy harvesting source and the power grid respectively. The consumed energy from the energy harvesting source cannot exceed the battery level. Hence, the energy causality constraint is given by:
\begin{equation}
 \label{eq:7}
X^h(i) \leq B(i).
\end{equation}
The battery level update is expressed as:
\begin{equation}
 \label{eq:8}
   B(i+1)=  \min \left(B_{\max}, B(i)- X^h(i) + E(i)\right).
\end{equation}
We consider that the consumed grid energy at frame $i$ is weighted by a factor $W_{i}$~\cite{mme}. Hence, the grid energy cost which is considered in this paper as an objective function, is expressed as:
\begin{equation}
\label{eq:9}
\Delta= \sum_{i=1}^{L} W_{i} X^g(i).
\end{equation}
In Fig.~\ref{fig_new1}, we illustrate the proposed hybrid energy LoRa framework.

\section{Problem Formulation}
The objective is to minimize the grid energy cost while using the available harvested energy. Each device requires a minimum received SNR to be satisfied. The energy management should take advantage of the grid's power weight variations by consuming more grid power when the associated weight is low while storing the renewable energy for future use, and by consuming less grid power when its weight is high. Moreover, the channels and SFs may be optimally assigned to the devices. Only 6 LDs per channel can be served simultaneously at each frame. The aim will be to schedule the devices with the minimal grid energy cost while making use of the available harvested energy and ensuring a minimum SNR to each scheduled LD. The main problem can be formulated as:

\begin{equation}
\label{eq:10}
\begin{aligned}
& \underset{ \underset{k=1,\ldots,K,m=1,\ldots,M,i=1,\ldots,L} {\{\chi_{k,m}(i),\alpha_k(i),p_k(i) \}}  } {\min} \sum_{i=1}^{L} W_{i} X^g(i) \\
& \text{subject to} \\
& (\ref{eq:10}.a): \gamma_{k,m}(i) \geq \chi_{k,m}(i) \gamma_{th},~\forall k=1,\ldots,K,i=1,\ldots,L,\\
& (\ref{eq:10}.b): \sum_{i=1}^{l} X^h(i) \leq \sum_{i=1}^{l} E(i),~\forall l=1,\ldots,L\\
& (\ref{eq:10}.c): \sum_{i=1}^{l} E(i) - \sum_{i=1}^{l-1} X^h(i) \leq B_{max},~\forall l=2,\ldots,L,
\end{aligned}
\end{equation}
\begin{equation}
\nonumber
\begin{aligned}
& (\ref{eq:10}.d): X^g(i)+X^h(i)=E_c+  \sum_{k=1}^{K} p_{k}(i) 2^{\alpha_k(i)} T,\\
& ~~~~~~~~~~~ \forall i=,\ldots,L,\\
& (\ref{eq:10}.e): \sum_{m=1}^{M} \chi_{k,m}(i) \leq 1,~\forall k=1,\ldots,K,i=1\ldots L,\\
& (\ref{eq:10}.f): \sum_{k=1}^{K} \chi_{k,m}(i) = 6,~\forall m=1,\ldots,M,i=1\ldots L,\\
& (\ref{eq:10}.g): \alpha_k(i) \neq \alpha_p(i),~\forall k,p \in \psi_m(i),k\neq p,\\
& ~~~~~~~~~~~ m=1,\ldots,M,,i=1,\ldots,L,\\
& (\ref{eq:10}.h): p_k(i) \geq 0,~\forall k=1,\ldots,K,i=1,\ldots,L,\\
& (\ref{eq:10}.i): \alpha_k(i) \in \Gamma,~\forall k=1,\ldots,K,i=1,\ldots,L\\
& (\ref{eq:10}.j): \chi_{k,m}(i) \in \{0,1\},~\forall k=1,\ldots,K,m=1,\ldots,M,\\
&  ~~~~~~~~~~~ i=1,\ldots,L.
\end{aligned}
\end{equation}
Constraint $(\ref{eq:10}.a)$ ensures a minimum received SNR, denoted $\gamma_{th}$, to each LD. Constraint $(\ref{eq:10}.b)$ is related to the energy causality, i.e. the consumed harvested energy cannot exceed the available energy at the battery. Additionally, constraint $(\ref{eq:10}.c)$ implies that the harvested energy at the current frame cannot exceed the maximum battery capacity. Constraint $(\ref{eq:10}.d)$ specifies that the required consumed energy is drawn from the grid and the energy harvesting source. Constraint $(\ref{eq:10}.e)$ imposes that only one channel at most can be assigned to each LD. Constraint $(\ref{eq:10}.f)$ specifies the maximal number of LDs within same channel, which cannot exceed the number of SFs. Constraint $(\ref{eq:10}.g)$ imposes that the LDs within same channel should be assigned different SFs. Constraint $(\ref{eq:10}.h)$ ensures the non-negativity of the allocated amounts of power. Constraint $(\ref{eq:10}.i)$ specifies the set of available SFs. Finally,  constraint $(\ref{eq:10}.j)$ specifies the channel assignment index.
\section{Optimal Offline Resource Management}
In this section, the optimal channel and SF assignment, as well as the energy management are investigated. The main formulated problem (\ref{eq:10}) is a mixed integer non-linear program because of its combinatorial nature and the non-linearity of the constraints. Meanwhile, to solve (\ref{eq:10}), the problem may be decoupled into two sub-problems. Since the goal is to minimize the total consumed energy at each frame, the channels and SFs may be optimally assigned among the LDs at each frame. Hence, the optimal total required consumed energy for network operation at each frame could be determined. Next, the optimal energy drawn from the energy harvesting source over time may be optimally derived based on the grid's wight.\\
First, the required transmit power to meet the SNR constraint of LD $k$ using channel $m$ at frame $i$ following (\ref{eq:4}) is given by:
\begin{equation}
\label{eq:11}
p_k(i) = \chi_{k,m}(i) \frac{\gamma_{th} \sigma_m^2}{\mid \mathbf{g}_{k,m}(i) \mid^2}.
\end{equation}  
Hence, the required consumed energy at frame $i$ is given by:
\begin{equation}
\label{eq:12}
X(i) = E_c + \sum_{k=1}^{K} \left( \sum_{m=1}^{M} \frac{  \gamma_{th} \sigma_m^2 }{\mid \mathbf{g}_{k,m}(i) \mid^2} \chi_{k,m}(i) \right) 2^{\alpha_k(i)}T.
\end{equation}  
Hence, replacing $p_k(i)$ and $X(i)$ by their expression in (\ref{eq:11}) and (\ref{eq:12}), the channel and SF assignment problem at frame $i$ can be formulated as: 
\begin{equation}
\label{eq:13}
\begin{aligned}
& \underset{ \underset{k=1,\ldots,K,m=1,\ldots,M} {\{\chi_{k,m}(i),\alpha_k(i) \}}  } {\min} \sum_{k=1}^{K} \left( \sum_{m=1}^{M} \frac{  \gamma_{th} \sigma_m^2 }{\mid \mathbf{g}_{k,m}(i) \mid^2} \chi_{k,m}(i) \right) 2^{\alpha_k(i)} \\
& \text{subject to} \\
& (\ref{eq:13}.a): \sum_{m=1}^{M} \chi_{k,m}(i) \leq 1,~\forall k=1,\ldots,K,\\
& (\ref{eq:13}.b): \sum_{k=1}^{K} \chi_{k,m}(i) = 6,~\forall m=1,\ldots,M,\\
& (\ref{eq:13}.c): \alpha_k(i) \neq \alpha_p(i),~\forall k,p \in \psi_m(i),k\neq p, m=1,\ldots,M,\\
& (\ref{eq:13}.d): \alpha_k(i) \in \Gamma,~\forall k=1,\ldots,K,\\
& (\ref{eq:13}.e): \chi_{k,m}(i) \in \{0,1\},~\forall k=1,\ldots,K,m=1,\ldots,M.
\end{aligned}
\end{equation}
The problem (\ref{eq:13}) is combinatorial and non-linear; and thus is a non-linear integer problem. Consequently, the problem is NP-hard~\cite{np} and could be solved by brute-force search with exponential complexity growth.\\
The required consumed energy for the network operation could be determined at each frame after deriving the optimal channel and SF assignment. Since the goal is to minimize the grid energy cost, the energy drawn from the energy harvesting source may be optimally managed over time. Hence, the energy management problem can be formulated as:
\begin{equation}
\label{eq:14}
\begin{aligned}
& \underset{ \underset{i=1,\ldots,L} {\{X^h(i) \}}  } {\min} - \sum_{i=1}^{L} W_{i} X^h(i) \\
& \text{subject to} \\
& (\ref{eq:14}.a): \sum_{i=1}^{l} X^h(i) \leq \sum_{i=1}^{l} E(i),~\forall l=1,\ldots,L,\\
& (\ref{eq:14}.b): \sum_{i=1}^{l} E(i) - \sum_{i=1}^{l-1} X^h(i) \leq B_{max},~\forall l=2,\ldots,L,
\end{aligned}
\end{equation}
\begin{equation}
\nonumber
\begin{aligned}
& (\ref{eq:14}.c): X^h(i) \leq X(i),~\forall i=1,\ldots,L,\\
& (\ref{eq:14}.d): X^h(i) \geq 0,~\forall i=1,\ldots,L.
\end{aligned}
\end{equation}
The objective function and the constraints of problem (\ref{eq:14}) are clearly linear. Hence, the optimal energy management is obtained by solving a linear program using interior-point method implemented in numerical tools such as CVX~\cite{cvx}.

\section{Online Resource Management}
In this section, online resource management is investigated for both scenarios uncorrelated and time-correlated channels by proposing low complexity heuristic algorithms. The LG is assumed to know the channel coefficients, the harvested energy and the grid's weight only at the current frame $i$.
\subsection{Heuristic Approach}
\subsubsection{Optimal SF Assignment}
The optimal SF assignment can be derived using the following theorem:\\
\textbf{Theorem~1.} Let consider $K$ LDs, where their coefficients $v_k$ verify $v_1<v_2<\ldots<v_K$. The optimal SF assignment that minimizes the objective function $\sum_{k=1}^{K} v_k 2^{\alpha_k}$ is given by assigning the lowest SF $\alpha^{\min}$ to the LD with biggest coefficient $v_k$ until assigning the biggest SF $\alpha^{\max}$ to the LD with lowest coefficient $v_k$.\\
\begin{proof}
Let consider two LDs with $v_1<v_2$. We have
\begin{equation}
\label{eq:15}
\begin{aligned}
v_1 2^7 + v_2 2^8 - (v_1 2^8 + v_2 2^7) & = - v_1 2^7 + v_2 2^ 7\\	
                                        & = 2^7 (v_2 - v_1) > 0.
\end{aligned}
\end{equation}
Hence, the optimal SF assignment is given by assigning 8 to LD 1 and 7 to LD 2. This relation can be extended recursively to the case of $K$ LDs.
\end{proof}

\subsubsection{Uncorrelated Channel}
Let consider a quasi-static Gaussian independent and identically distributed (i.i.d.) slow fading channel. A heuristic low complexity algorithm is proposed to solve channel and SF assignment, and energy management problem in LoRa network powered by energy harvesting. Only $N M$ LDs can be scheduled at each frame. The proposed algorithm starts by scheduling the LDs one by one based on their channel coefficient. The best channel is assigned to the LD with the highest channel coefficient modulus in order to save the required consumed energy. This procedure is repeated until all the channels become full.\\
Next, the SF assignment are performed following \textbf{Theorem~1} by assigning the lowest SF $\alpha^{\min}$ to the user with biggest coefficient $p_k$ until assigning the biggest SF $\alpha^{\max}$ to the LD with lowest coefficient $p_k$. Finally, after assigning the channels and SFs, the required consumed energy could be computed. The proposed algorithm  uses the maximum available harvested energy at the battery at each frame. The proposed Heuristic Uncorrelated Resource Management Algorithm (HURMA) is described in \textbf{Algorithm~1}.\\
The computational complexity of HURMA is derived as follows. For the SF assignment, an array with $N$ elements is sorted $M$ times in the \textbf{for} loop with a complexity order $O(M N \log(N))$. The channel assignment in the \textbf{while} loop has a complexity order equal to $O(M N \log(M N))$ because it is similar to sort an array with $M N$ elements. Hence, the computational complexity of HURMA is given by:
\begin{equation}
\begin{aligned}
\label{eq:16}
C^{\text{HURMA}}&= O(M N \log( N) + M N \log(M N)) \\
&=O(M N \log(M N)).
\end{aligned}
\end{equation}
Hence, the proposed low complexity algorithm can be executed in polynomial time.

\begin{algorithm}[h]
	\begin{algorithmic}
		\caption{Heuristic Uncorrelated Resource Management Algorithm (HURMA)}
		\State $B(1) \gets E(1)$, // \textit{battery initialization}
		\For{$i=1:L$}
		\State $\chi_{k,m}(i) \gets 0$, // \textit{initialization}
		\State  $\bm{c} \gets \bm{0}_{K \times 1}$, // \textit{initialize the vector that contains the channel index for each LD}
		\State  $\bm{c}_f \gets \bm{0}_{ M \times 1}$, // \textit{initialize the vector that indicates the number of scheduled LD for each channel}
		\State $\bm{V}$ is a matrix that contains the modulus of the channels coefficients for all LDs
		\While{$\sum_{m=1}^M \bm{c}_f[m] \neq M N$}
		\State $(k_{\max},m_{\max}) \gets \argmax \bm{V}$, // \textit{select the LD with higher channel coefficient}
		\If{$\bm{c}_f[m_{\max}] < N$}
		\State $\bm{c}[k_{\max}]  \gets m_{\max}$, // \textit{assign channel $m_{\max}$ to LD $k_{\max}$}
		\State $\bm{c}_f[m_{\max}] \gets \bm{c}_f[m_{\max}] +1$
		\State $\chi_{k_{\max},m_{\max}}(i) \gets 1$, // \textit{LD selection}
		\State $\bm{V}[k_{\max},:]  \gets \bm{0}_{1 \times M} $
		\Else
		\State $\bm{V}[:,m_{\max}]  \gets \bm{0}_{K \times 1} $
		\EndIf	
		\EndWhile
		\For{$m=1:M$}
		\State $\bm{d}_m \gets$ indices of LDs assigned to $m$
		\State  $v_n \gets \frac{ \sigma_m^2}{\mid \mathbf{g}_{\bm{d}_m(n),m}(i) \mid^2},n=1,\ldots,N$
		\State $\bm{s}_m \gets$ $\bm{d}_m$ sorted in ascending order based on $p_n$
		\State  $\alpha_{\bm{s}_m(n)}(i) \gets 13-n,n=1,\ldots,N $, // \textit{SF assignment}    
		\EndFor
		\State $X(i) \gets E_c + \sum_{k=1}^{K} \left( \sum_{m=1}^{M} \frac{  \gamma_{th} \sigma_m^2 }{\mid \mathbf{g}_{k,m}(i) \mid^2} \chi_{k,m}(i) \right) 2^{\alpha_k(i)}T$, // \textit{compute the required consumed energy} 
		\If{$X(i) \leq B(i) $}
		\State $X^h(i) \gets X(i)$
		\State $X^g(i) \gets 0$
		\Else
		\State $X^h(i) \gets B(i)$ 
		\State $X^g(i) \gets X(i)-X^h(i)$   
		\EndIf
		\State $B(i) \gets \min \left(B_{\max}, B(i)- X^h(i) + E(i)\right)$, // \textit{battery update}
		\EndFor
	\end{algorithmic}
\end{algorithm}

\subsubsection{Time-correlated Channel}
Let us consider a time-correlated channel which is modeled by Gilbert Elliot channel model~\cite{elliot}. The state of the channel at frame $i$ is modeled as a one-dimensional Markov chain with two states: a good state denoted by $\bm{G}$, and a bad state denoted by $\bm{B}$. Channel transitions occur at the beginning of each frame. The transition probabilities are given by:
\begin{equation}
\label{eq:333333}
\mathbb{P}\left[h_{k,m}(i)=\bm{G} \mid h_{k,m}(i-1)=\bm{G}\right]=\lambda_{1},
\end{equation}
and 
\begin{equation}
\label{eq:444444}
\mathbb{P}\left[h_{k,m}(i)=\bm{G} \mid h_{k,m}(i-1)=\bm{B}\right]=\lambda_{0}.
\end{equation}
A second heuristic resource management algorithm is proposed taking into account the channel correlation. The channel assignment is performed as follows. First, the $N M$ LDs with highest path loss coefficient are selected. Then, the proposed algorithm starts by assigning the channels to the LDs one by one based on their path loss coefficient starting from the most distant LD. The corresponding LD chooses the best available channel, i.e., the one with highest channel coefficient modulus. We start by the most distant LD by assigning it the best channel in order to compensate their SNR and save the required consumed energy. This procedure is repeated until all the channels become full. Next, the SF assignment and energy management are performed similar to the algorithm HURMA. The proposed Heuristic Correlated Resource Management Algorithm (HCRMA) is described in \textbf{Algorithm~2}.\\
The computational complexity of HCRMA is derived as follows. The SF assignment for HCRMA is similar to HURMA and is done with a complexity order $O(M N \log(N))$. The channel assignment in the \textbf{for} loop has a complexity in the order $O(M^2 N)$. The sort of the array $\bm{w}$ is done with a complexity $O(K \log(K))$. Hence, the computational complexity of HCRMA is given by:
\begin{equation}
\label{eq:17}
C^{\text{HCRMA}}= O(M N \log(N)+ M^2 N +K \log(K)).
\end{equation}
The proposed low complexity algorithm can be executed in polynomial time.

\begin{algorithm}[h]
	\begin{algorithmic}[0]
		\caption{Heuristic Correlated Resource Management Algorithm (HCRMA)}
		\State $B(1) \gets E(1)$, // \textit{battery initialization}
		\For{$i=1:L$}
		\State $\chi_{k,m}(i) \gets 0$, // \textit{initialization}
		\State  $\bm{c} \gets \bm{0}_{K \times 1}$, // \textit{initialize the vector that contains the channel index for each LD}
		\State  $\bm{c}_f \gets \bm{0}_{ M \times 1}$, // \textit{initialize the vector that indicates the number of scheduled LD for each channel}
		\State $\bm{w}$ contains the ascending sort of the indices of the LDs based on path loss
			\For{$k=K-M N+1:K$}
		\State $k_{\max} \gets \bm{w}[k]$
		\State $m_{\max} \gets$ the index of the channel with highest coefficient		 for LD $k_{\max}$
		\algstore{long}
	\end{algorithmic}
\end{algorithm}
\begin{algorithm}[h]
\addtocounter{algorithm}{-1}
		\caption{Heuristic Correlated Resource Management Algorithm (HCRMA)}
	\begin{algorithmic}[0]
	\algrestore{long}
		 \State $\bm{c}[k_{\max}]  \gets m_{\max}$, // \textit{assign channel $m_{\max}$ to LD $k_{\max}$}
		 \State $\chi_{k_{\max},m_{\max}}(i) \gets 1$, // \textit{LD selection}
		 \State $\bm{c}_f[m_{\max}] \gets \bm{c}_f[m_{\max}] +1$
		\If{$\bm{c}_f[m_{\max}] = N$}
	    \State remove channel $m_{\max}$ from the set of available channels
		\EndIf	
		\EndFor
		\For{$m=1:M$}
		\State $\bm{d}_m \gets$ indices of LDs assigned to $m$
		\State  $v_n \gets \frac{ \sigma_m^2}{\mid \mathbf{g}_{\bm{d}_m(n),m}(i) \mid^2},n=1,\ldots,N$
		\State $\bm{s}_m \gets$ $\bm{d}_m$ sorted in ascending order based on $p_n$
		\State  $\alpha_{\bm{s}_m(n)}(i) \gets 13-n,n=1,\ldots,N $, // \textit{SF assignment}    
		\EndFor
		\State $X(i) \gets E_c + \sum_{k=1}^{K} \left( \sum_{m=1}^{M} \frac{  \gamma_{th} \sigma_m^2 }{\mid \mathbf{g}_{k,m}(i) \mid^2} \chi_{k,m}(i) \right) 2^{\alpha_k(i)}T$, // \textit{compute the required consumed energy}
		\If{$X(i) \leq B(i) $}
		\State $X^h(i) \gets X(i)$
		\State $X^g(i) \gets 0$
		\Else
		\State $X^h(i) \gets B(i)$ 
		\State $X^g(i) \gets X(i)-X^h(i)$   
		\EndIf
		\State $B(i) \gets \min \left(B_{\max}, B(i)- X^h(i) + E(i)\right)$, // \textit{battery update}
		\EndFor
	\end{algorithmic}
\end{algorithm}
\subsection{Reinforcement Learning Approach}
 \begin{figure*}[h]
\centering
	\includegraphics[scale=0.75]{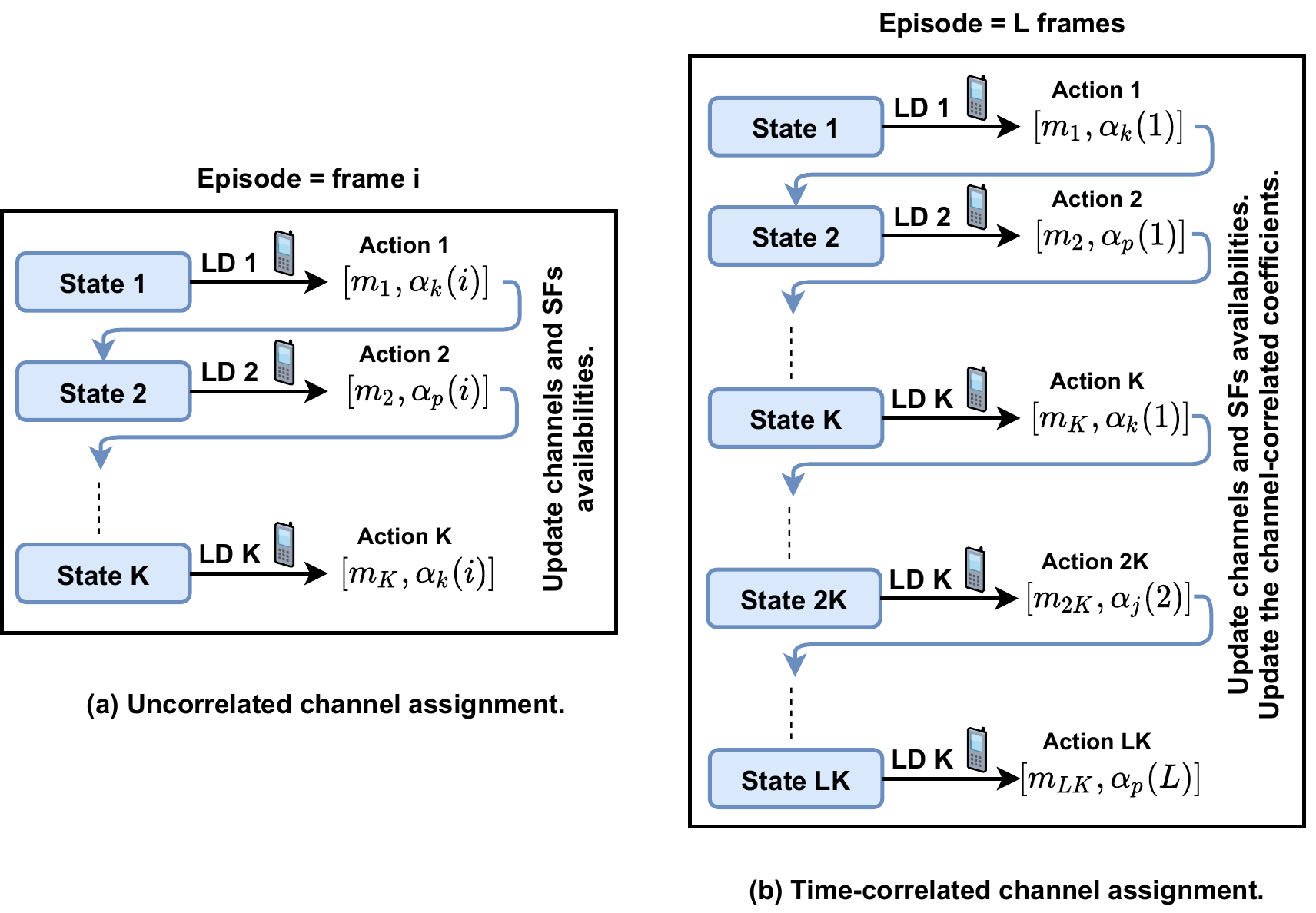}
	\caption{Episode design of uncorrelated and time-correlated channel assignments.}
	\label{episode}
\end{figure*}
The problem in (\ref{eq:10}) is NP-hard, which makes deriving the optimal solution extremely complex and time consuming. Moreover,
in the optimization, we suppose that we have a full overview about
the future harvested energy, which is not realistic. On the other hand, the online heuristic presents a low-complexity solution. However, allocation decisions are taken greedily, without any insight about the harvested energy in the next frames. Recently, reinforcement learning techniques have become highly adopted for applications characterised by dynamic and complex problem spaces. More specifically, RL is considered as one of the important paradigms of machine learning, in addition to supervised and unsupervised
learning~\cite{RL,DQN}. The advantage of this technique is that it approaches the optimal solution by interacting with the environment parameters (e.g., the harvested energy and the energy prices), learning the statistical distributions of these features, and determining the most efficient policy that takes the actions based on the current status of the system and the learned insights about the future. Therefore, the reinforcement learning can be the most adequate solution to reduce the complexity of the NP-hard problem, while taking sub-optimal decisions owing to the knowledge learned about the system.

On these bases, we propose to investigate an online resource management solution based on RL. In the previous section, we decomposed the optimal solution (\ref{eq:10}) into two sub-problems, namely channel assignment and energy management problems. Accordingly, we design two RL systems, responsible to solve the above-mentioned sub-problems. In particular, in each time step of the resource management process, the RL agents should make a decision on the channels to allocate, the assigned SFs, and the optimal amount of energy to draw  from  the  energy  harvesting  source. The agents ultimately aim to minimize the grid energy cost, while respecting the battery capacity and the constraints on the available resources. During the learning process, the energy harvesting system receives rewards and experiences penalties for each allocation decision it makes until it reaches the convergence to the optimal policy. The defined sub-problems can be abstracted as two Markov Decision Process (MDP) frameworks; each one is presented by the five-tuple $(S, A, P, R, \gamma)$~\cite{RL}. $S$ presents the set of states for each framework, $A$ denotes the set of potential actions, $P$ is the state transition probability, $R$ is defined as the immediate reward gained for each action, and $\gamma$ denotes the discount factor. These elements are discussed for both sub-problems, in the following sub-sections.
\subsubsection{Channel assignment problem}
 \begin{figure*}[h]
\centering
	\includegraphics[scale=0.75]{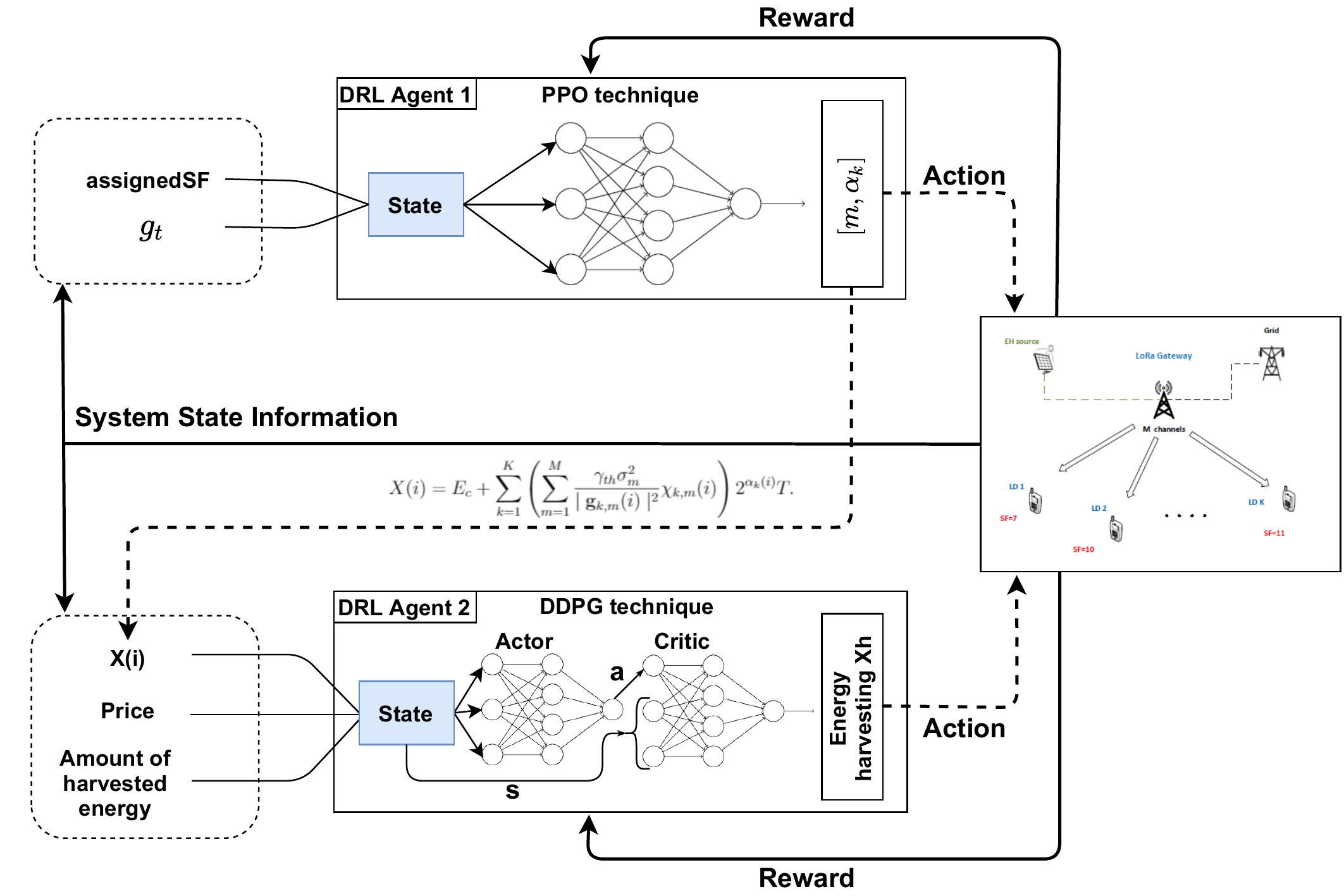}
	\caption{Deep reinforcement learning architecture.}
	\label{System}
\end{figure*}
\begin{itemize}[noitemsep,topsep=0pt,leftmargin=*]
  \item \textbf{MDP environment design:} In our paper context, the MDP environment represents the LoRa wireless network with which the agent interacts. We design this environment to receive the action $A_t$ generated by the agent at a step $t$, assign a reward $R_t$, and introduce the next state $S_{t+1}$. The finite sequence of such steps is called an episode. By experiencing various episodes, the agent is trained to learn from historical actions and their associated rewards. Therefore, in the first RL framework, we define each step as the channel and spreading factor allocation to one of the LoRa devices, as illustrated in the sub-problem (\ref{eq:13}). We underline that different episodes of experiences are independent, which means that at the beginning of each episode, the cumulative reward is initiated to 0. Therefore, since the channels can be time-correlated or totally uncorrelated, the episode is defined for each network configuration differently. More specifically, if channels are uncorrelated among different $L$ frames, we define each episode as one frame $i$ where channels and SFs are assigned to the existing LDs. This way, the episode length is equal to the number of devices $K$. In case the channels are correlated, the resource management of all frames should be accomplished in one episode, implying that the episode length is equal to $K \times L$. The illustration of both episode designs is presented in Fig. \ref{episode}. It is worth mentioning that the agent does not have an overview of the environment design. Instead, the optimal policy $\Pi: S \to A $ is built by observing the surrounding environment, selecting actions, and gaining rewards.
  \item \textbf{States and actions:} The set of states $S$ is composed of all possible environment circumstances and conditions at each step. We set $S=\{assignedSF(i), g_{t}(i)\}$, where $assignedSF(i)$ is an $M \times 6$ matrix of assigned SFs in the frame $i$. $assignedSF(i)_{m,j}$ is equal to 1, if the spreading factor $j$ of the channel $m$ is assigned to one of the LDs, 0 otherwise. $assignedSF(i)$ is initiated to a null matrix at each new frame. $g_t$ is the matrix of the channels' coefficients between the  gateway  and  the LD related to the current step $t$ and frame $i$. Depending on the system state and based on the policy $\Pi$, the agent takes an action $A_t$. This action includes selecting the appropriate channel $m$ and the related spreading factor $\alpha_k$. Thus, the action can be expressed as $A_t=[m,\alpha_k]$. Note that $m \in \{0..M\}$ and $\alpha_k \in \Gamma$, as indicated by the constraints (\ref{eq:13}.d) and (\ref{eq:13}.e). We note that $m$ equal to 0 means that the LD will not be assigned a channel. The constraint (\ref{eq:13}.a) is respected by design as we assign at most one channel $m$ to each LD at each step. Once the decision is made, the LD related to the current step is served through the chosen resources. After each step, the $assignedSF$ matrix is updated according to the predicted action. Subsequently, the new matrix is fed as an input to the next step as illustrated in Fig. \ref{episode}.
  \item \textbf{Reward function:} The reward function is formulated to match the high-level objective of the optimization problem (\ref{eq:13}), aiming at minimizing the required energy at each frame, while respecting the availability of channels and spreading factors. To ensure respecting such constraints, rewards and penalties are assigned. Specifically, when a state $S_t$ is received, a decision $A_t$ should be taken while meeting the following requirements:
  \begin{equation}\label{reward}
  \small
  \begin{aligned}
\left \{
  \begin{array}{lr}
    \text{$C_1$:} \quad \sum_{j=(i.K+1)}^{t}(A_j(1)==A_t(1)) \leq 6 &
    \text{constraint (\ref{eq:13}. b)}\\
    \text{$C_2$:}\quad assignedSF_{A_t(1),A_t(2)}(i)=0 & \text{constraint (\ref{eq:13}. c)} 
  \end{array}
\right.
   \end{aligned}
\end{equation}
$C_1$ indicates that the maximal number of LDs within the chosen channel cannot exceed the number of SFs, which matches the constraint (\ref{eq:13}.b). $C_2$ shows that the chosen spreading factor associated to the allocated channel is not assigned to another LD, which is equivalent to the constraint (\ref{eq:13}.c). We note that $i$ denotes the current frame, which is equal to $\floor*{\frac{t-1}{K}}$. To this end, we define the immediate reward as follows:
\begin{equation}\label{rewardFunction}
    \begin{aligned}
R_t=C_1*C_2- \frac{  \gamma_{th} \sigma_{A_t(1)}^2 }{\mid \mathbf{g}_{t,A_t(1)}(i) \mid^2} 2^{A_t(2)}.
    \end{aligned}
\end{equation}
If one of the constraints $C_1$ or $C_2$ is not respected ($C_1=0$ or $C_2=0$), invalid situations can occur. Hence, we attribute 0 as a reward. The maximum direct reward is only received, when all requirements are met. In this way, the RL system tries to  meet the scenarios that respect the constraints, in order to maximize the received bonuses and avoid null rewards. Additionally to the reward assigned for respecting the constraints of the LoRa system, the RL agent is charged for inaccurate resource allocation. Since the  performance of the network is  quantified by minimizing the required energy at each frame $i$, we use this indicator to define the penalty added to the reward function, as shown in Eq. (\ref{rewardFunction}). Accordingly, the RL agent selects optimal allocations by maximizing the cumulative rewards and minimizing the penalties.
\item \textbf{Agent design:} The goal of the agent is to learn how to minimize the required energy throughout different episodes. To achieve this goal in the long run, the agent needs to build an optimal policy $\Pi$ that maximizes the future expected reward approximated by the action-value function $Q^{\Pi}$ expressed by the classical Bellman equation:
\begin{equation}\label{eq:Q}
\begin{aligned}
Q^{\Pi}(s,a)=\mathbb{E}[\sum_{k=0}^{T}\gamma^k R_{t+k}|\Pi,S_{t}=s,A_{t}=a],
\end{aligned}
\end{equation}
where $0 \leq \gamma \leq 1$ denotes the discount parameter, that reflects the importance of the direct reward compared to long-term reward received at the end of the episode. Setting $\gamma$ to be small implies that the agent is designed to be shortsighted, and only the step rewards are considered. A bigger $\gamma$ indicates that the agent is farseeing and gives higher weights to future rewards. The Q-value $Q(S_t,A_t)$ serves to assess the accuracy of the decision $A_t$ for a given state $S_t$. After determining the optimal policy, the agent selects its actions as follows:
\begin{equation}\label{eq:Q}
\begin{aligned}
\Pi^*(s) = arg max_a Q^{\Pi}(s,a).
\end{aligned}
\end{equation}
Since our LoRa system is dynamic and the action space is dimensional and depends on the number of channels and spreading factors, it is challenging to save all Q-values in a Q-table and use traditional RL methods \cite{QLearning}. Consequently, we propose to adopt Deep Reinforcement Learning (DRL) using DNNs to approximate the action-value function. DRL approaches can be classified into two categories: the value-based and policy-based methods. Particularly, the value-based method adopts deep learning to estimate the value function (e.g., DQN \cite{DQN}), whereas the policy-based DRL uses DNNs for approximating the parameterized policy (e.g., REINFORCE \cite{reinforce}). We opt for the latter approach as it achieves better performance for stochastic policies. This approach works by computing an estimator of the policy gradient:
\begin{equation}\label{Policy_gradient}
\begin{aligned}
\nabla L(\theta)=\mathbb{E}[\nabla_{\theta} log \Pi(S_t|A_t,\theta) \hat{E^s_t}],
\end{aligned}
\end{equation}
where $\Pi$ is the parametrized policy, $\theta$ represents the weight of the DNN, and $\hat{E^s_t}$ is the function estimator at the step $t$. $\hat{E^s_t}$ is calculated as follows:
\begin{equation}\label{estimator}
\begin{aligned}
\hat{E^s_t}=\sum_{i=0}^{\infty} (\gamma \lambda)^i \delta^s_{t+i},
\end{aligned}
\end{equation}
\begin{equation}
\begin{aligned}
\delta^s_{t}=R_t+\gamma V(S_{t+1},\theta)-V(S_t,\theta),
\end{aligned}
\end{equation}
where $\lambda$ is used to adjust the bias–variance trade-off and $V(S_t,\theta)$ is the state-value defined as the expected return  when being in state $S_t$ and parametrized by $\theta$.

To enhance the exploration ability of the policy-based approaches, model-free and on-policy learning is introduced, where the learning is based on historical actions and the current policy, without any knowledge about the environment. One of the most known on-policy algorithms proposed by OpenAI is Proximal Policy Optimization (PPO) \cite{ppo}. The objective function of PPO is presented by:
\begin{equation}\label{ppo}
\begin{aligned}
L^{CLIP}(\theta)=\mathbb{E}[(p_t(\theta)\hat{E^s_t},clip(p_t(\theta),1-\epsilon,1+\epsilon)\hat{E^s_t})],
\end{aligned}
\end{equation}
where the policy probability ratio $p_t(\theta)$ is defined as:
\begin{equation}\label{g_t}
\begin{aligned}
p_t(\theta)=\frac{\Pi(A_t|S_t,\theta)}{\Pi(A_t|S_t,\theta_{old})}.
\end{aligned}
\end{equation}
The clip function is responsible to constraint $p_t$ between $(1-\epsilon)$ and $(1+\epsilon)$, which prevents the system from moving outside this interval. In this way, the PPO objective is limited to the lower bound of the unclipped part. Owing to these advantages, we adopt the PPO method to design our channel assignment RL system.
\item \textbf{DRL algorithm:} To stabilize the training and ensure the convergence of the learning process, multiple steps must be followed, which we illustrate in Algorithm \ref{PPO}. First, two networks are initialized with the same weights, from which two PPO policies are established. The training process starts by generating samples from the policy with fixed parameters $\theta_{old}$, for different episodes. These samples are saved in a replay memory D (lines 16-18). More specifically, the agent receives the set of observations from the environment and takes an action using the policy $\Pi_{\theta_{old}}$ aiming at being highly rewarded (lines 11-15). After each episode, the estimator is computed and a random mini-batch from D is sampled. Next, the main policy $\Pi_{\theta}$ is updated by calculating the gradient of $\theta$ for each sample of the batch (lines 21-25). This mechanism, known as experience replay, is very important to learn from past experience and to reach the convergence and stabilization of the learning. In the end, we synchronize both policies by replacing $\theta_{old}$ with $\theta$ (line 29).
\end{itemize}
\begin{algorithm}[!h]
\caption{Channel assignment RL system}
\label{PPO}
\begin{algorithmic}[1]
\State \textbf{ Initialization:}
\State - Randomly set the parameters $\theta$ of the DNN to get $\Pi_{\theta}$.
\State - Set the sampling policy $\Pi_{\theta_{old}}$ with $\theta_{old} \leftarrow \theta$.
\If{\text{channel is uncorrelated}}
\State F=1
\Else{ F=L}
\EndIf
\State \textbf{ DRL Learning:}
\For{each episode $e$}
\State $i=0$
\For{each frame $f=1..F$}
\State $assignedSF(f)=zeros(M,6)$
\For{each step $t=i+1..i+K$}
\State $S_t=\{assignedSF(f), g_t\}$
\State Select $A_t$ based on $\Pi_{\theta_{old}}$
\State $R_t=C_1*C_2- \frac{  \gamma_{th} \sigma_{A_t(1)}^2 }{\mid \mathbf{g}_{t,A_t(1)}(i) \mid^2} 2^{A_t(2)}$
\State Update $assignedSF(f)$
\State  Observe $R_t$ and the next state $S_{t+1}$.
\State  Save $(S_t,A_t,R_t,S_{t+1})$  in  the experience
\State memory $D$.
\EndFor
\State $i=i+K$
\EndFor
\State - Compute the estimator $\hat{E_t^s}$ according to (\ref{estimator}).
\State - Sample a mini-batch of $(S_j,A_j,R_j,S_{j+1})$
\State  from the memory $D$.
\State - Update $\theta$ by maximizing the objective  
\State function (\ref{ppo}) and using the sampled data.
\State - Update the old policy: $\theta_{old} \leftarrow \theta$
\EndFor
			\end{algorithmic}
\end{algorithm}
\subsubsection{Energy management problem}
\begin{itemize}[noitemsep,topsep=0pt,leftmargin=*]
\item \textbf{MDP environment design:} Similarly to the first problem, the second RL agent responsible for the energy management interacts with the LoRa wireless network. However, in this problem, an episode denotes the management of energy among L frames, while each step throughout the episode denotes the allocation of the energy $X^h$ at a frame $t$. Therefore, the episode length will be equal to $L$.
\item \textbf{States and actions:} At each time step $t$, the set of states received by the agent comprises the
following components: $E(t)$ which is the amount of harvested energy, $X(t)$ which denotes the required energy at the frame $t$, and $W_t$  presenting the energy cost. We remind that $X(t)$ is derived from the decision of the first RL agent, as depicted in Fig. \ref{System}. The system state $S_t$ is, therefore, a vector defined as: $S_t=\{E(t),X(t),W_t\}$. Next, by observing the state, the agent decides the amount of energy $X^h(t) \geq 0$ to be harvested from the energy source. Finally, after each step, the set of states $S_{t+1}$ is generated.
\item \textbf{Reward function:} The reward function in the second RL system should match the objective of the optimization problem (\ref{eq:14}) aiming to minimize the grid energy cost, which is equivalent to maximizing $W_t X_h$. In addition, this function should guarantee that the action $A_t= X^h(t)$ is taken while meeting the following requirements:
\begin{equation}\label{reward2}
    \begin{aligned}
\left\{
  \begin{array}{lr}
    \text{$C_3$:} \quad  \sum_{i=1}^t A_i\leq \sum_{i=1}^t E(i) & \text{constraint (\ref{eq:14}.a)}\\
    \text{$C_4$:}\quad \sum_{i=1}^{t-1} E(i)-\sum_{i=1}^t A_i\leq B_{max}  & \text{constraint (\ref{eq:14}.b)}\\ 
    \text{$C_5$:}\quad A_t \leq X(t) & \text{constraint (\ref{eq:14}.c)}
  \end{array}
\right.
    \end{aligned}
\end{equation}
The constraints $C_3$, $C_4$, and $C_5$ in (\ref{reward2}) are set to match the equations (14a), (14b), and (14c), respectively. Accordingly, the immediate reward is defined as follows:
\begin{equation}\label{rewardFunction2}
    \begin{aligned}
R_t=C_3*C_4*C_5+(W_t*A_t)* \mathbbm{1}_{(C_3*C_4*C_5=1)} 
   \end{aligned}
\end{equation}
As in the first RL model, the non-respect of one of the constraints (set to $C_3*C_4*C_5=-1$ in this section) incurs invalid scenarios. This means that maximizing the rewards involves respecting the constraints to avoid the negative penalties. Moreover, in the second model, the high-level goal is to maximize the harvested energy. Thus, the weight of energy $W_t*A_t$ is added to the reward function, if all constraints are respected. In this way, maximizing the cumulative rewards throughout the episodes implies maximizing the amount of harvested energy and consequently reducing the usage of grid energy.
\item \textbf{Agent design:} The energy management RL model is characterized by a dynamic state space and a non-discrete action space. Based on these factors, we select a Deep Deterministic Policy Gradient (DDPG)-based method to optimize the decision-making policy. DDPG is an actor-critic, off-policy, and model-free algorithm known for its high performance on continuous action and state spaces \cite{ddpg}. The actor-critic algorithms are generally composed of a policy and an action-value functions, where the policy function plays the role of the actor that takes decisions and interacts with the environment, whereas the action-value function is called a critic that is responsible to evaluate the performance of the actor.
\item \textbf{DRL algorithm:} Two deep neural networks are adopted to build the approximation functions of the DDPG algorithm. The first DNN is used to train the actor and it is defined by the policy function $\mu(s|\theta^{\mu})$ and the weights $\theta^{\mu}$. The second network, which corresponds to the critic, is described by the action-value function $Q(s,a|\theta^Q)$ with the related weights equal to $\theta^Q$. The actor-critic networks are illustrated in Fig. \ref{System}. As done in the first system,  steps to follow are illustrated in Algorithm \ref{DDPG}. First, copies of the main networks, namely target networks, are created with the same NN parameters, i.e., $\theta'^{\mu}=\theta^{\mu}$ and $\theta'^Q=\theta^Q$ (lines 2-4). 
Next, the learning process is performed by generating episodes of experiences repeatedly (lines 6-31) and storing the generated MDP tuples in a replay buffer $D$ (line 14). To ensure that the agent discovers different possible actions, an exploration policy is constructed by adding a noise sample through a noise process $N^s$. The selected action under the state $S_t$ is defined by $A_t=\mu(s|\theta^{\mu})+N^s_t$  (line 9). Once the decision is made, the direct reward is assigned and a new state $S_{t+1}$ is given. 

To improve the critic policy, the agent samples each  step a random mini-batch from the replay buffer $D$ (lines 17-18) and calculates the target values $y(j)$ for each sample using the critic target network (lines 19-20). Meanwhile, the actor target network generates an action $\mu_{target}(S_{j+1})$ and feeds it to the critic target network. The DDPG critic policy is presented by the classical Bellman equation illustrated in (\ref{eq:Q}).
This critic network is updated by reducing the loss illustrated in (line 20) and expressed as:
\begin{equation}\label{loss}
    \begin{aligned}
L=\frac{1}{B} \sum_j (y(j)-Q(S_j,A_j|\theta^Q))^2.
    \end{aligned}
\end{equation}

\begin{algorithm}[h]
\caption{Energy management RL system}
\label{DDPG}
\begin{algorithmic}[1]
\State \textbf{ Initialization:}
\State - Randomly set the parameters $\theta^{\mu}$ of the actor network 
\State \quad and $\theta^Q$ of the critic networks.
\State - Set weights of target
networks: $\theta'^{\mu} \leftarrow \theta^{\mu}$ and $\theta'^Q \leftarrow \theta^Q$.
\State \textbf{ DDPG Learning:}
\For{each episode $e$}
\For{each step $t=1..L$}
\State $S_t=\{E(t),X(t),W_t\}$
\State Select $A_t=\mu(S_t|\theta^\mu)+N^s_t$
\If{$C_3*C_4*C_5=1$}
\State $R_t=W_t*A_t$
\Else { $R_t=-1$}
\EndIf
\State - Observe $R_t$ and the next state $S_{t+1}$.
\State - Save $(S_t,A_t,R_t,S_{t+1})$  in an experience \State memory $D$.
\State - Sample a mini-batch of $(S_j,A_j,R_j,S_{j+1})$ of  
\State  size $B$ from the memory $D$.
\State - Find target Q-value $y(j)$ from target  Q-network:
\State  {\small $y(j)=R_j+\gamma Q_{target}(S_{j+1},\mu_{target}(S_{j+1}|\theta'^{\mu})|\theta'^Q)$}
\State - Update the weights $\theta^Q$ of the critic network by
\State reducing the loss:
\State $L=\frac{1}{B} \sum_j (y(j)-Q(S_j,A_j|\theta^Q))^2$
\State - Update the weights $\theta^{\mu}$ of the actor network using
\State the gradient policy:
\State {\small $\nabla_{\theta^{\mu}}J \simeq \frac{1}{B}\sum_j \nabla_{a} Q(s,a|\theta^Q)|_{s=S_j,a=\mu(S_j)}\nabla_{\theta^{\mu}}\mu (s|\theta^{\mu})|_{S_j}$}
\State - Update the target networks:
\State $\theta'^{Q} \leftarrow \rho \theta^Q +(1-\rho) \theta'^Q$
\State $\theta'^{\mu} \leftarrow \rho \theta^{\mu} +(1-\rho) \theta'^{\mu}$
\EndFor
\EndFor
			\end{algorithmic}
\end{algorithm}

On the other hand, the actor network $\mu$ is updated using the gradient policy:
\begin{equation}\label{GP}
    \begin{aligned}
\nabla_{\theta^{\mu}}J \simeq E[ \nabla_{\theta^{\mu}} Q(s,a|\theta^Q)|_{s=S_t, a=\mu(S_t|\theta^{\mu})}].
    \end{aligned}
\end{equation}
Similarly to the target network, the main actor generates an action $\mu(S_j)$ and inputs it to the critic network. However, in this process, the action and weights gradients, namely $\nabla_{\theta^{\mu}} Q(S_j,A_j|\theta^Q)$ and $\nabla_{\theta^{\mu}}\mu (S_j|\theta^{\mu})$, are calculated using the
automatic differentiation technique. These gradients allow the approximation of the global policy gradient:
 \begin{equation}\label{}
    \begin{aligned}
\nabla_{\theta^{\mu}}J \simeq \frac{1}{B}\sum_j \nabla_{a} Q(s,a|\theta^Q)|_{s=S_j,a=\mu(S_j)}\nabla_{\theta^{\mu}}\mu (s|\theta^{\mu})|_{S_j}.
    \end{aligned}
\end{equation}
Finally, the target networks are softly updated using a small rate $\rho$, to always represent the most recent learning (lines 28-29): 
\begin{equation}\label{update}
   \begin{aligned}
\theta'^{Q} \leftarrow \rho \theta^Q +(1-\rho) \theta'^Q, \quad
\theta'^{\mu} \leftarrow \rho \theta^{\mu} +(1-\rho) \theta'^{\mu}.
    \end{aligned}
\end{equation}
\end{itemize}

The theoretical complexity of our DRL resource management approach in hybrid energy LoRa wireless networks is based on the complexity of both RL systems. More specifically, the complexity of Algorithm \ref{PPO} to accomplish one episode is determined by the loops between lines 11 and 21, where the number of iterations is equal to $LK$. Meanwhile, the complexity of Algorithm \ref{DDPG} is determined by one loop performing $L$ iterations. Therefore, the complexity of the whole system can be expressed as $O(L(K+1))$. Furthermore, the complexity of the DNN decisions inside these loops depends on the neural network. In our case,  we adopt an MLP deep neural network comprising 2 layers of 64 neurons, which is a light-weight network. 

\section{Evaluation Results}
In this section, Monte Carlo and reinforcement learning simulations are done to evaluate the proposed resource management schemes in LoRa wireless networks by averaging up to 10000 realizations. Monte Carlo simulations are based on repeating random sampling to obtain numerical results. The underlying concept is to use randomness to solve problems having a probabilistic interpretation. In our work, we use Monte Carlo simulation to simulate the random variables that represent the channel, energy, and device distribution. The users are uniformly distributed within a circular cell. The grid energy consumption weights $W_{i}$ are randomly generated according to a standard uniform distribution. The simulation parameters used in this section are summarized in Table~\ref{tab1}.\\
\begin{table}[h]
	\centering
	\caption{Simulation Parameters.}
	\label{tab1}
	\begin{tabular}{|c|c|c|}
		\hline  \textbf{Symbol} & \textbf{Description} &  \textbf{Value}  \\
		\hline   $B_{max}$ & max battery capacity & 200 J \\
		\hline   $K$ & number of devices & 35 \\
		\hline   $L$ & number of frames & 50 \\
		\hline   $M$ & number of channels & 5 \\
		\hline       & path loss exponent & 3.7~\cite{elliot} \\
		\hline       & noise PSD & -174 dBm/Hz~\cite{elliot} \\
		\hline       & circuit power & 30 dBm~\cite{lte}\\
		\hline       & cell radius & 500 m \\
		\hline
	\end{tabular}
\end{table}
The RL algorithms are validated based to the parameters defined in Tables \ref{tab:PPO} and \ref{tab:DDPG}. These parameters are empirically adopted and we expect that, using the same values, similar architectures perform identically.
\begin{table}[h]
\centering
\caption{Hyper-parameters of the channel assignment RL system.}
\label{tab:PPO}
\begin{tabular}{|l|l|l|}
\hline
\textbf{Parameter} & \textbf{Description} & \textbf{Value} \\ \hline
$\gamma$ & Gamma & 0.99 \\ \hline
$\alpha$ & Learning rate & 0.0001 \\ \hline
Policy & DNN policy & MLP, 2 layers of 64 \\ \hline
 $\epsilon$ & cliprange & 0.2 \\ \hline
$\lambda$ & Adjusting factor & 0.01 \\ \hline
& PPO epoch & 4 \\ \hline
\end{tabular}
\end{table}
\begin{table}[h]
\centering
\caption{Hyper-parameters of the energy management RL system.}
\label{tab:DDPG}
\begin{tabular}{|l|l|l|}
\hline
\textbf{Parameter} & \textbf{Description} & \textbf{Value} \\ \hline
$\gamma$ & Gamma & 0.99 \\ \hline
$\alpha_a$ & \begin{tabular}[c]{@{}l@{}}Learning rate\\ of actor\end{tabular} & 0.0001 \\ \hline
$\alpha_c$ & \begin{tabular}[c]{@{}l@{}}Learning rate\\ of critic\end{tabular} & 0.001 \\ \hline
Policy & DNN policy & MLP, 2 layers of 64 \\ \hline
bz & Buffer size & 25000 \\ \hline
$N^s_t$ & \begin{tabular}[c]{@{}l@{}}Noise\\ parameter\end{tabular} & None \\ \hline
$\rho$ & soft update & 0.001 \\ \hline
\end{tabular}
\end{table}
\subsection{Uncorrelated Channel}
\begin{figure}[H]
	\centerline{\includegraphics [width=1.05\columnwidth]  {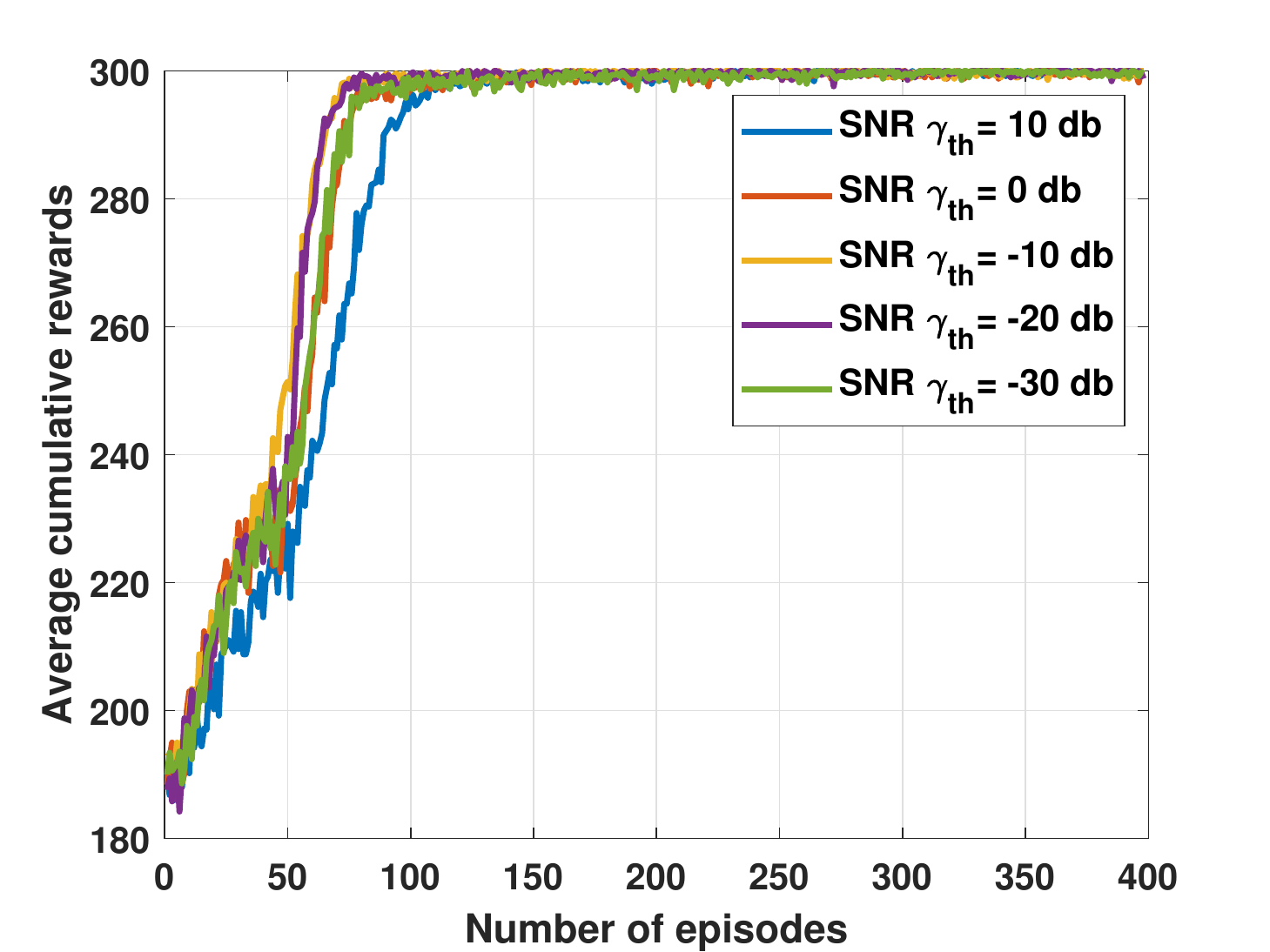}}
	\caption{Average cumulative rewards vs. training episodes of the channel allocation RL system ($M=2,K=6,B_{max}=10~\text{J},\Gamma=\{7,8,9\}$).}
	\label{convergence_uncorrelated}
\end{figure}

\begin{figure}[H]
	\centerline{\includegraphics [width=1.05\columnwidth]  {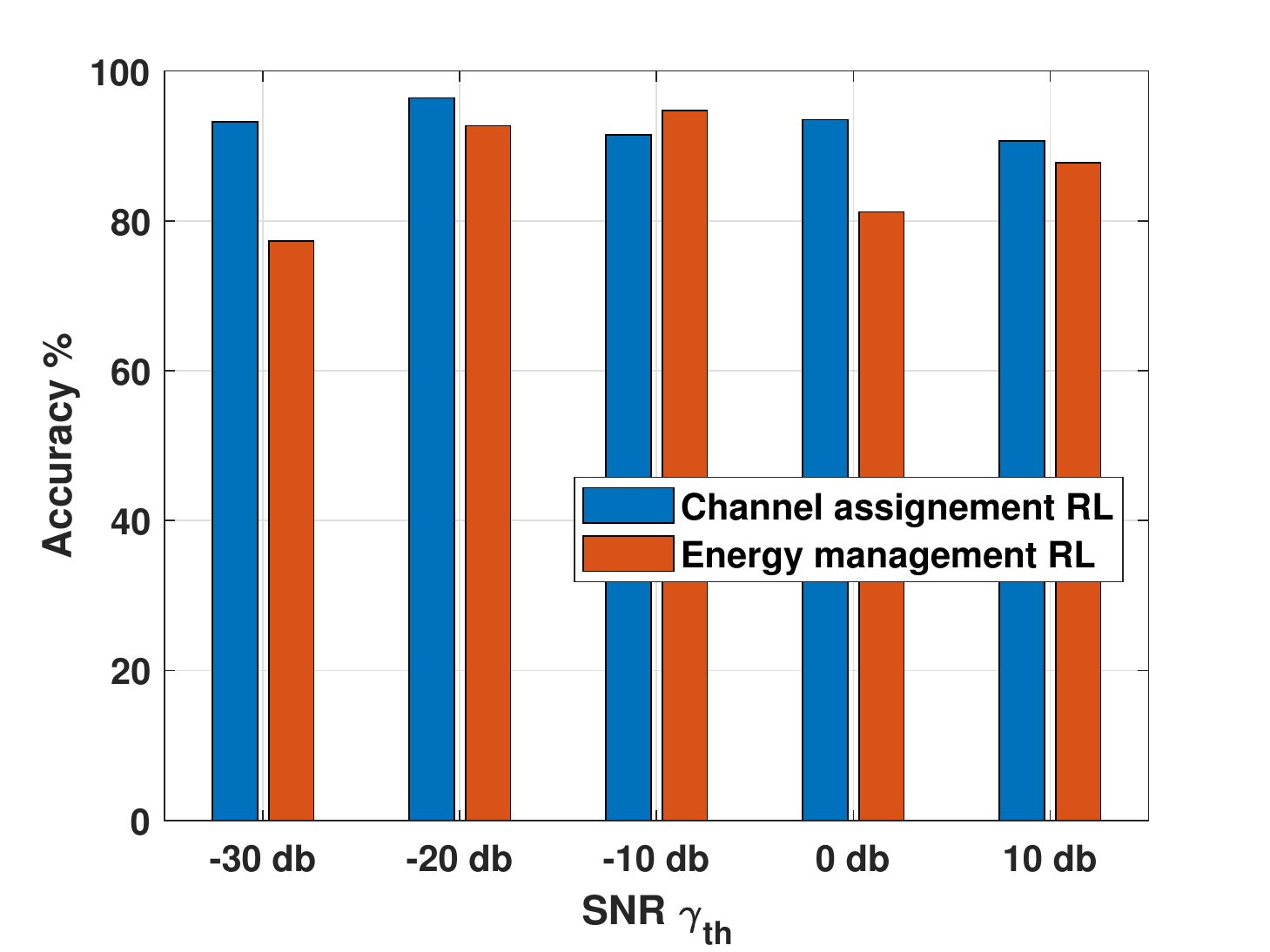}}
	\caption{Accuracy in terms of meeting constraints for both RL systems ($M=2,K=6,B_{max}=10~\text{J},\Gamma=\{7,8,9\}$).}
	\label{accuracy_uncorrelated}
\end{figure}

First, we investigate the performance of the resource management schemes in LoRa networks considering uncorrelated channels. Particularly, we will start by examining the performance of the RL approaches in terms of convergence and ability to respect the system constraints. 
Fig.~\ref{convergence_uncorrelated} illustrates the variation of the cumulative rewards over the training episodes, for the channel allocation RL system.
We note that the presented rewards are averaged over a window of 5 for all SNR levels to see the behavior of the system during training. In the beginning of the learning process, all decisions are taken randomly in order to have an initial estimation of the RL policy. At this stage, we can notice that the reward is low, which means that the constraints described in eq. (\ref{reward}) are not respected. However, as the number of episodes increases, the system starts to learn how to respect different constraints and assign optimal allocations. After the learning process, the stability is reached, which confirms the convergence of the channel allocation RL system.\\
Fig.~\ref{accuracy_uncorrelated} shows the accuracy of both RL systems, namely channel assignment and energy management. We define the accuracy of the RL as its ability to respect different constraints. More specifically, the accuracy is equal to the percentage of episodes where all system requirements are satisfied, after the convergence. We can see that the accuracy of both RL models is very high reaching 80\% and more for most of the SNR levels, which means more than 80\% of episodes respect the defined constraints (e.g. channels, SFs, and energy constraints).\\ 
\begin{figure}[!h]
	\centerline{\includegraphics [width=1.05\columnwidth]  {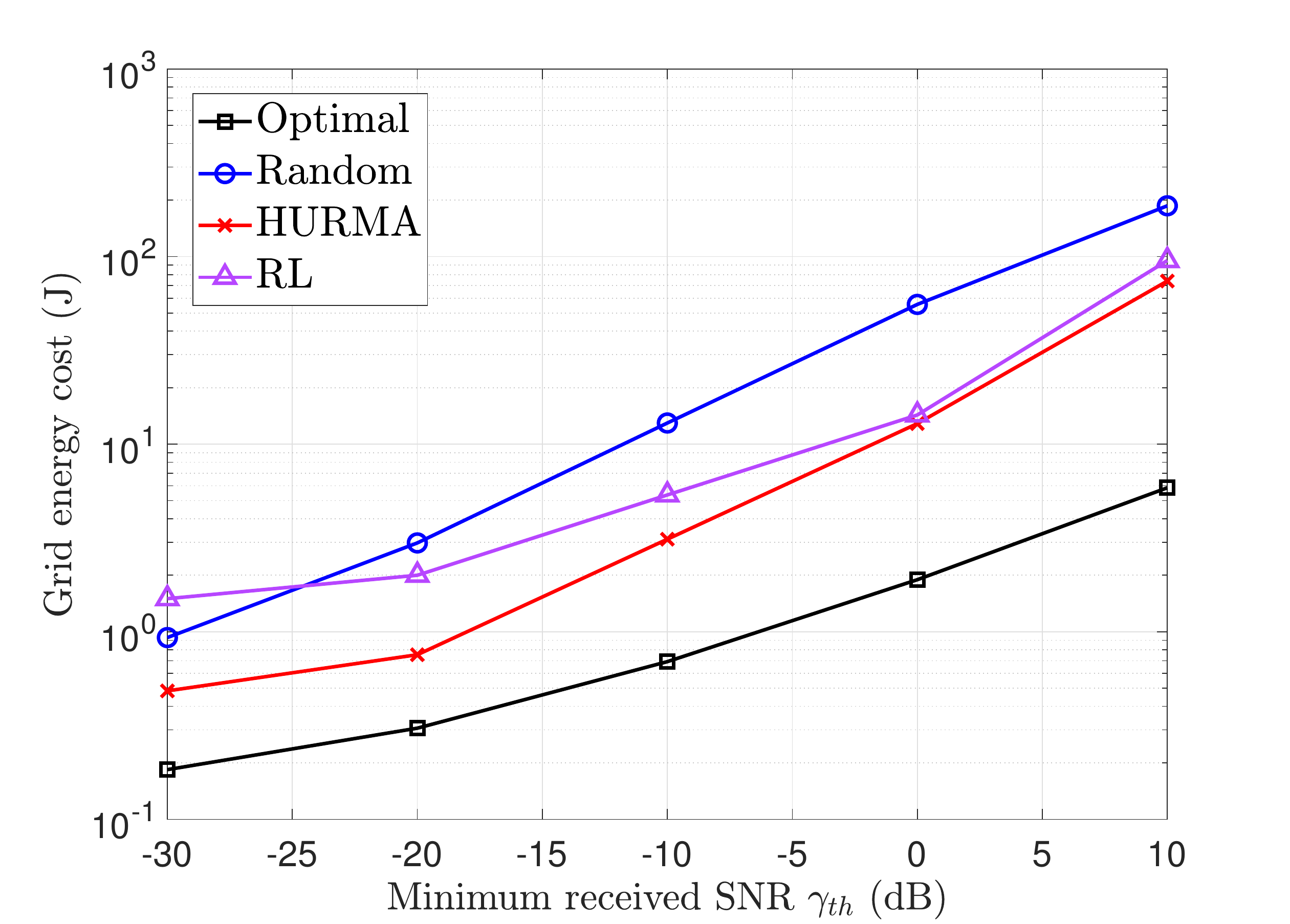}}
	\caption{Optimal grid energy cost versus SNR target ($M=2,K=6,B_{max}=10~\text{J},\Gamma=\{7,8,9\}$).}
	\label{fig1}
\end{figure}
Next, we will present the performance of the proposed HURMA algorithm and energy management RL system compared to the optimal solution. 
\begin{figure}[!h]
	\centerline{\includegraphics [width=1.05\columnwidth]  {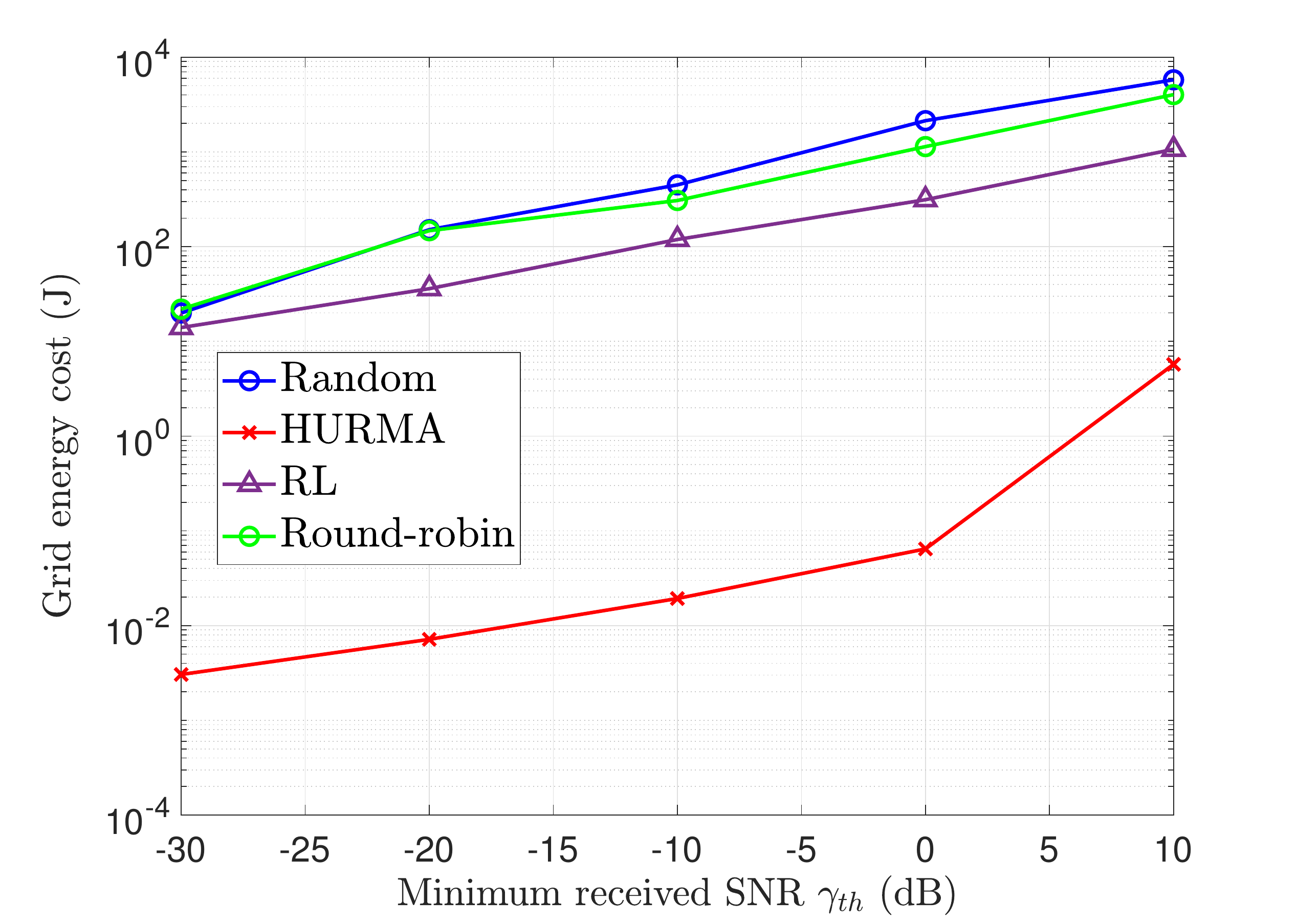}}
	\caption{Grid energy cost versus SNR target with heuristic resource management schemes ($M=5,K=35,B_{max}=200~\text{J},\Gamma=\{7,8,9,10,11,12\}$).}
	\label{fig2}
\end{figure}
\begin{figure}[!h]
	\centerline{\includegraphics [width=1.05\columnwidth]  {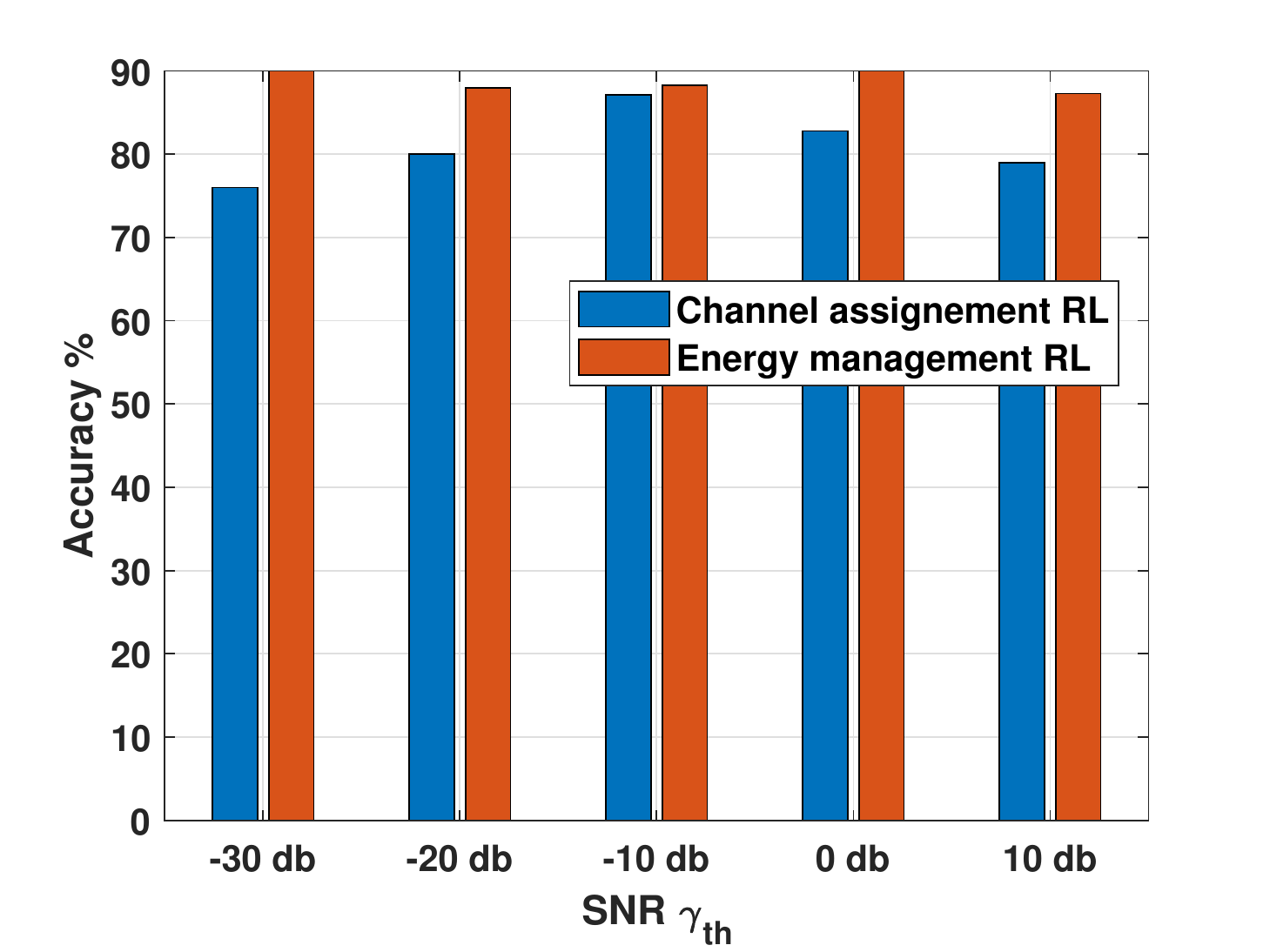}}
	\caption{Accuracy in terms of meeting constraints for both RL systems ($M=5,K=35,B_{max}=200~\text{J},\Gamma=\{7,8,9,10,11,12\}$).}
	\label{fig2_accuracy}
\end{figure}
Fig.~\ref{fig1} plots the Grid energy cost of the proposed low complexity algorithm HURMA and the RL system as a function of the SNR. This figure is drawn for only limited number of devices $K=6$, number of channels $M=2$ and a set of SFs $\Gamma=\{7,8,9\}$ due to the high complexity of the exhaustive search optimal algorithm. It is clear that the HURMA and the RL approach significantly outperform the random scheme (random channel and SF assignment) thanks to the adequate channel and SF assignment, which significantly saves the transmitted power. Also, they achieve a performance near to the optimal specifically in low SNR region in which LoRa operates. However, we can notice that the HURMA heuristic presents a better performance compared to the RL approach. This can be explained by the fact that no MDP process is set, when channels are uncorrelated. This means that the episode steps are independent and the state transition follows a uniform distribution. Therefore, the RL policy only learns how to respect the constraints to maximize the reward. Additionally, it captures the channels and SFs that have higher probabilities to minimize the required energy, owing to the experience memory $D$ storing past allocation decisions. 
\begin{figure}[!h]
	\centerline{\includegraphics [width=1.05\columnwidth]  {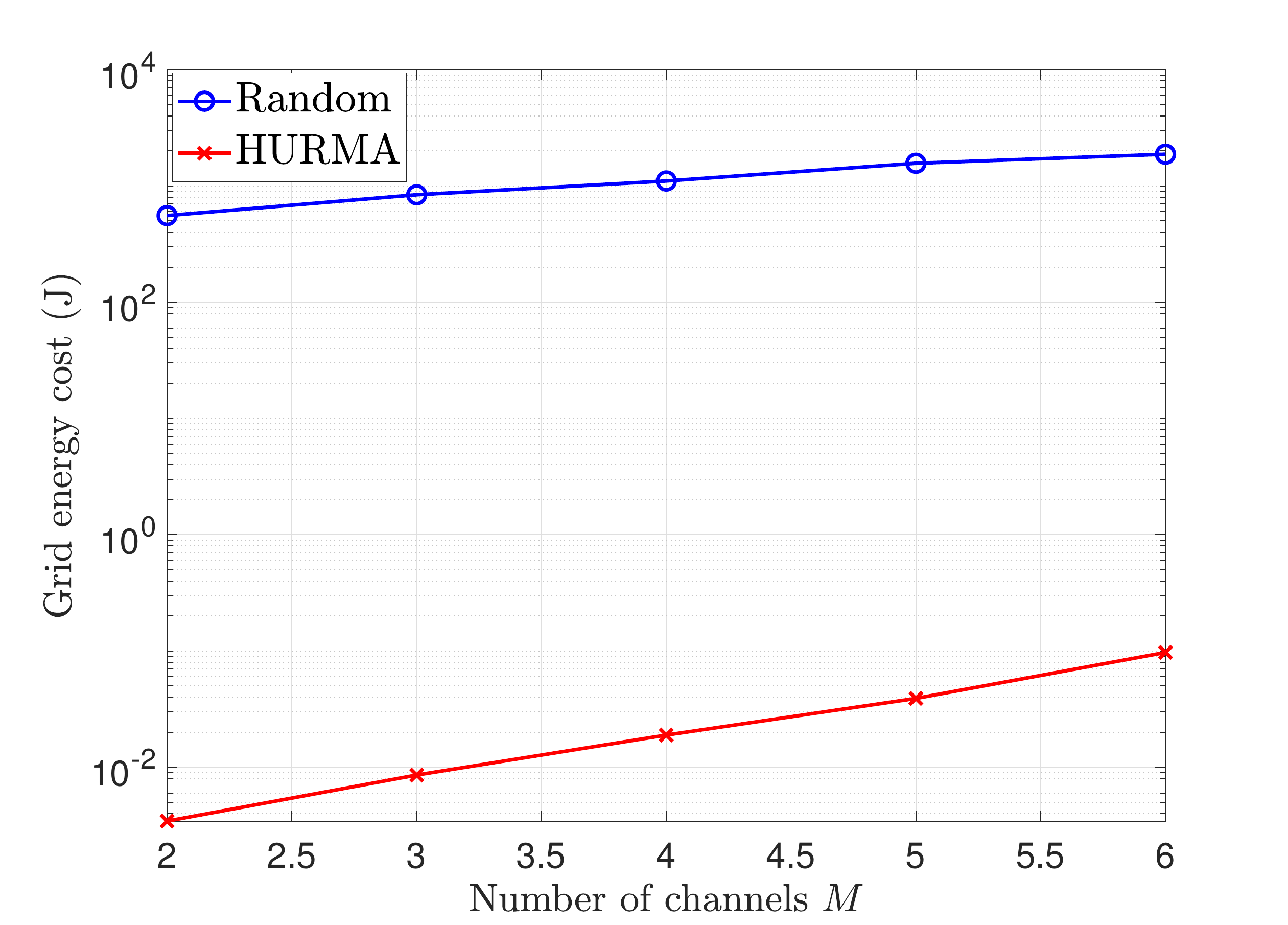}}
	\caption{Grid energy cost versus number of channels with heuristic resource management schemes ($K=40,B_{max}=200~\text{J},\Gamma=\{7,8,9,10,11,12\}, \gamma_{th}=0$ dB).}
	\label{fig3}
\end{figure}
\begin{figure}[!h]
	\centerline{\includegraphics [width=1.05\columnwidth]  {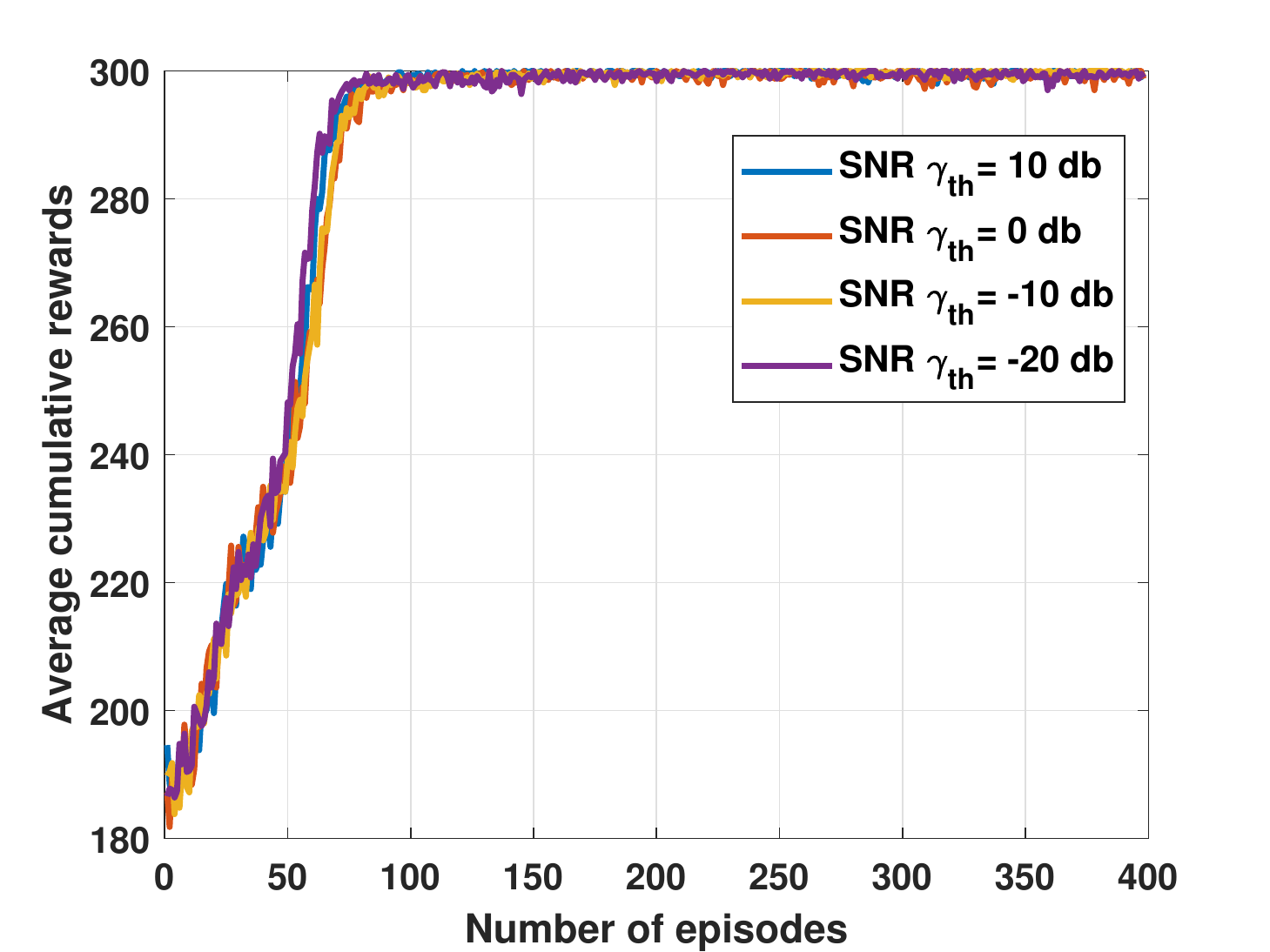}}
	\caption{Average cumulative rewards vs. training episodes of the channel allocation RL system ($M=2,K=6,B_{max}=10~\text{J},\Gamma=\{7,8,9\}$).}
	\label{convergence_correlated}
\end{figure}
To summarize, when channels are uncorrelated, HURMA outperforms the RL approaches due to its efficient energy management design. Still, the RL shows a good performance, while presenting a lower complexity. Particularly, our RL system uses an MLP predictive network composed of 2 layers of 64 neurons, which is a light-weight DNN model with a negligible complexity. Moreover, the edge of the RL over heuristic based approaches is its run-time ability to adapt to the environment changes (e.g., SNR, number of channels, and number of users.), thanks to its continual online learning.\\
In Fig.~\ref{fig2}, the performance of HURMA is shown as a function of the SNR for higher number of number of devices $K=35$, number of channels $M=5$ and a set of SFs $\Gamma=\{7,8,9,10,11,12\}$. It is clear that the proposed algorithm HURMA significantly saves grid energy cost compared to the RL approach and round-robin scheduling, which confirms the performance of the heuristic over the online learning when channels are uncorrelated. We note that the accuracy of both RL models leading to energy saving is verified in Fig.~\ref{fig2_accuracy}.\\ 
Fig.~\ref{fig3} shows the performance of the proposed scheme HURMA as a function of the number of channels $K$. The increase of the number of channels allows to schedules more devices, which increases the total grid energy cost. Moreover, HURMA performs very well and the performance gap with the random scheme keeps almost unchanged when the number of channels increases.
\subsection{Time-correlated Channel}
Now, the performance of the resource management schemes in LoRa networks is investigated considering time-correlated channels.
Fig.~\ref{convergence_correlated} depicts the average cumulative rewards of the channel allocation RL over the training episodes, smoothed over a window of 5. Similarly to the uncorrelated scenario, the system applies the trial and error process, until reaching the convergence phase.
\begin{figure}[h]
	\centerline{\includegraphics [width=1.05\columnwidth]  {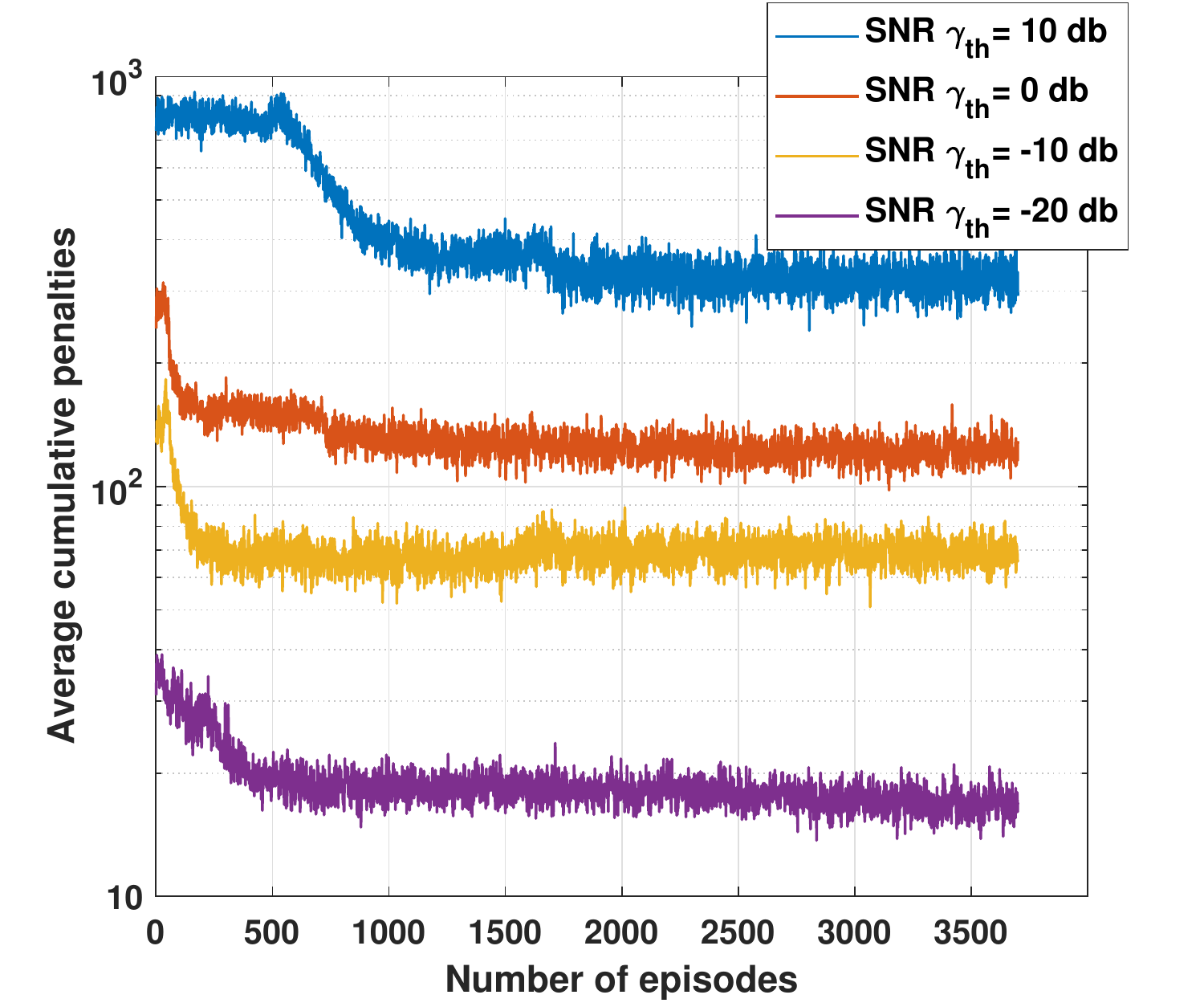}}
	\caption{Average cumulative penalties vs. training episodes of the channel allocation RL system ($M=2,K=6,B_{max}=10~\text{J},\Gamma=\{7,8,9\}$).}
	\label{penalties_correlated}
\end{figure}
\begin{figure*}[h]
\centering
	\mbox{
	\hspace{-0.4cm}
	     \subfigure[\label{ereq10}]{\includegraphics[scale=0.47]{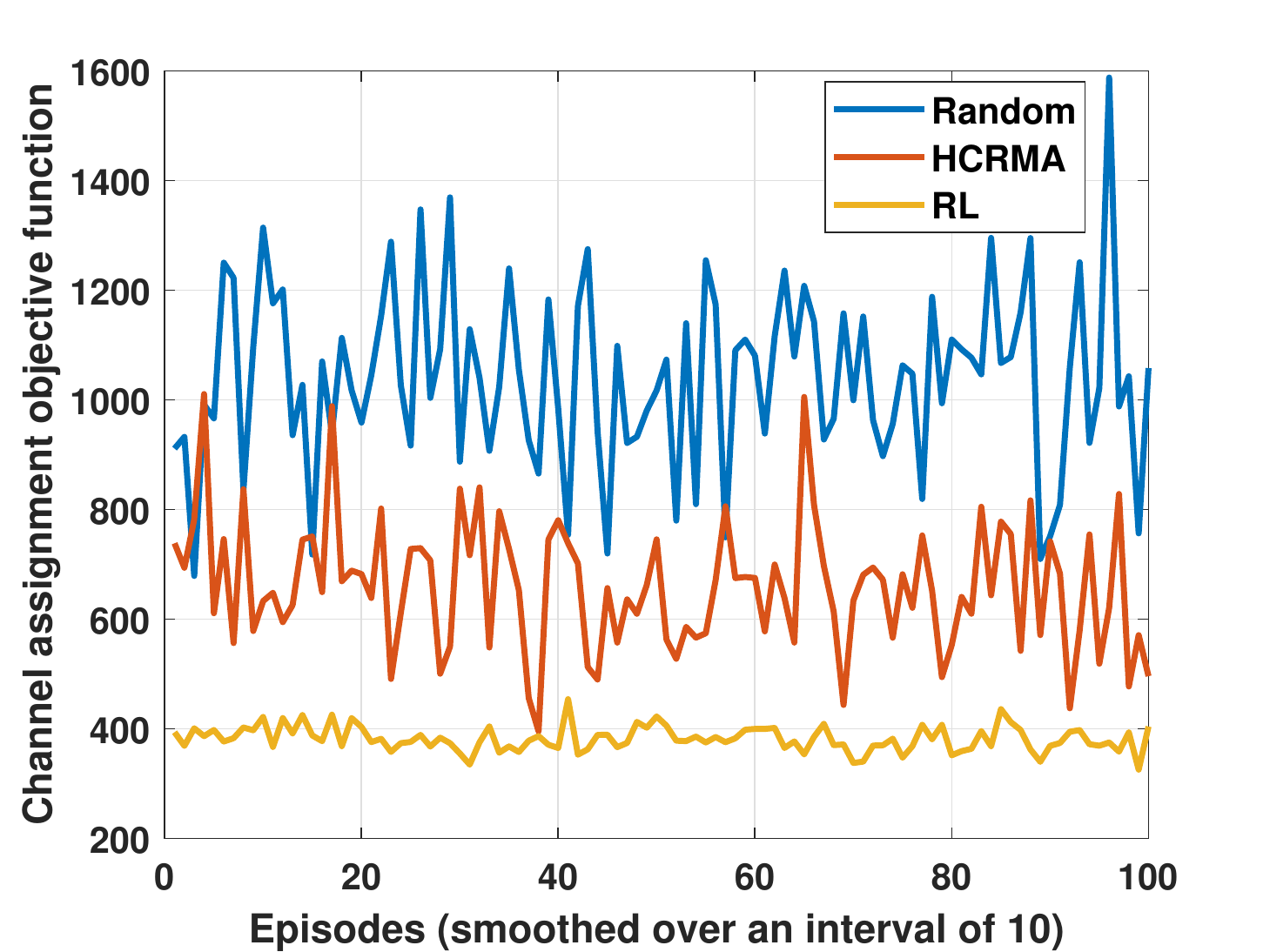}}
	     \hspace{-0.6cm}
	     \subfigure[\label{ereq0}]{\includegraphics[scale=0.47]{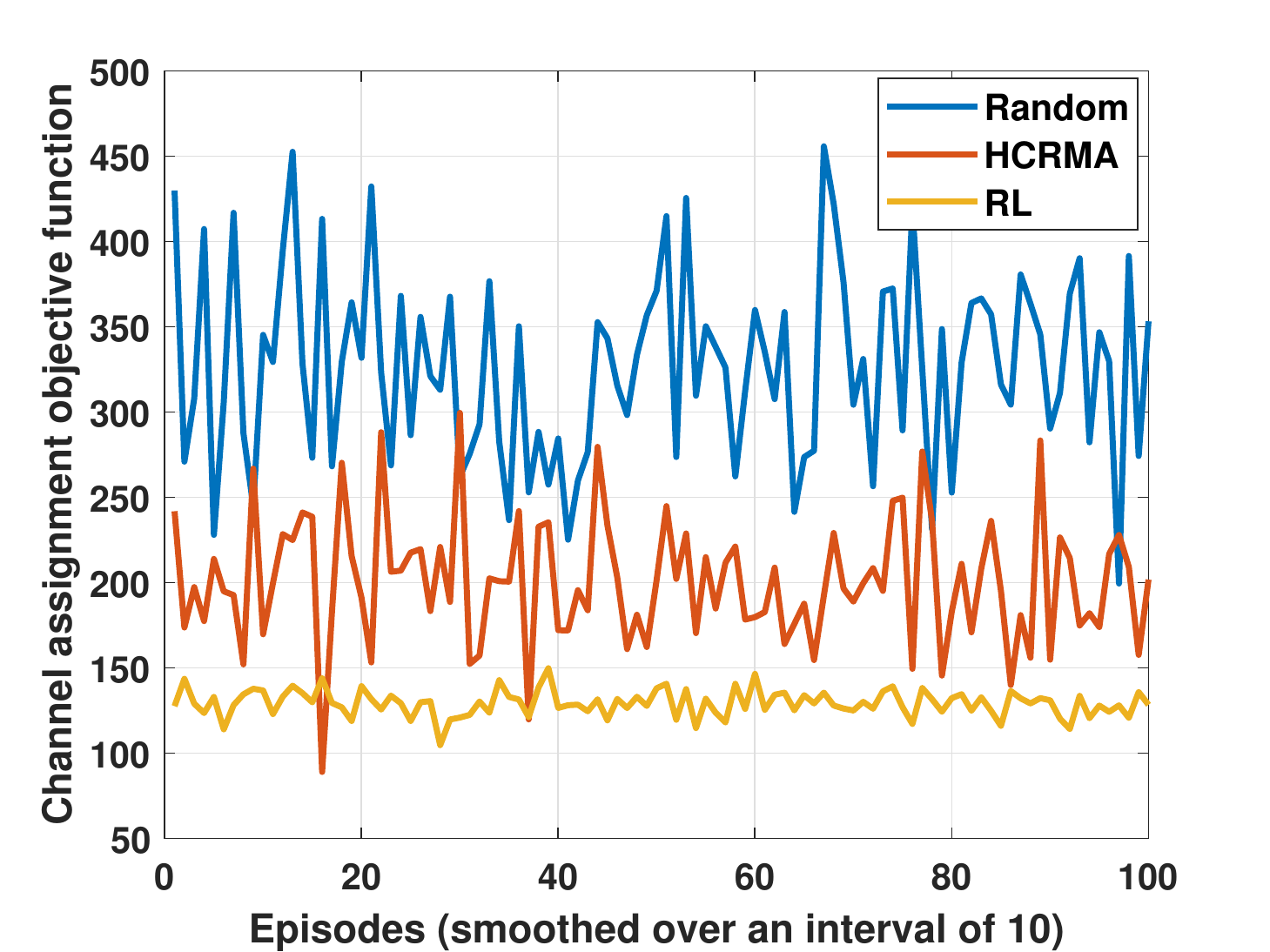}}
}\\
	\mbox{
	\hspace{-0.4cm}
	    \subfigure[\label{ereq-10}]{\includegraphics[scale=0.47]{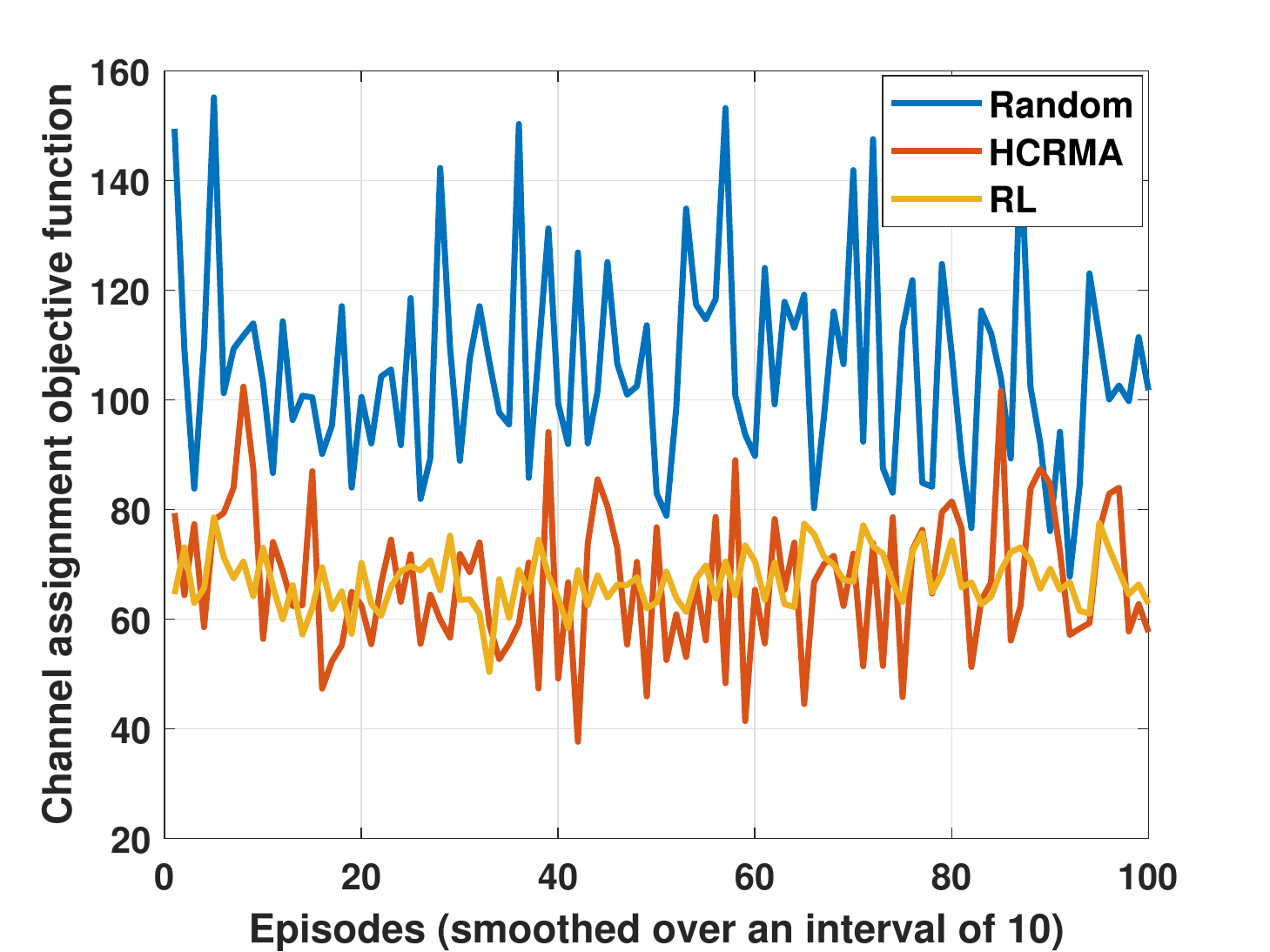}}
	    \hspace{-0.6cm}
        \subfigure[\label{ereq-20}]{\includegraphics[scale=0.47]{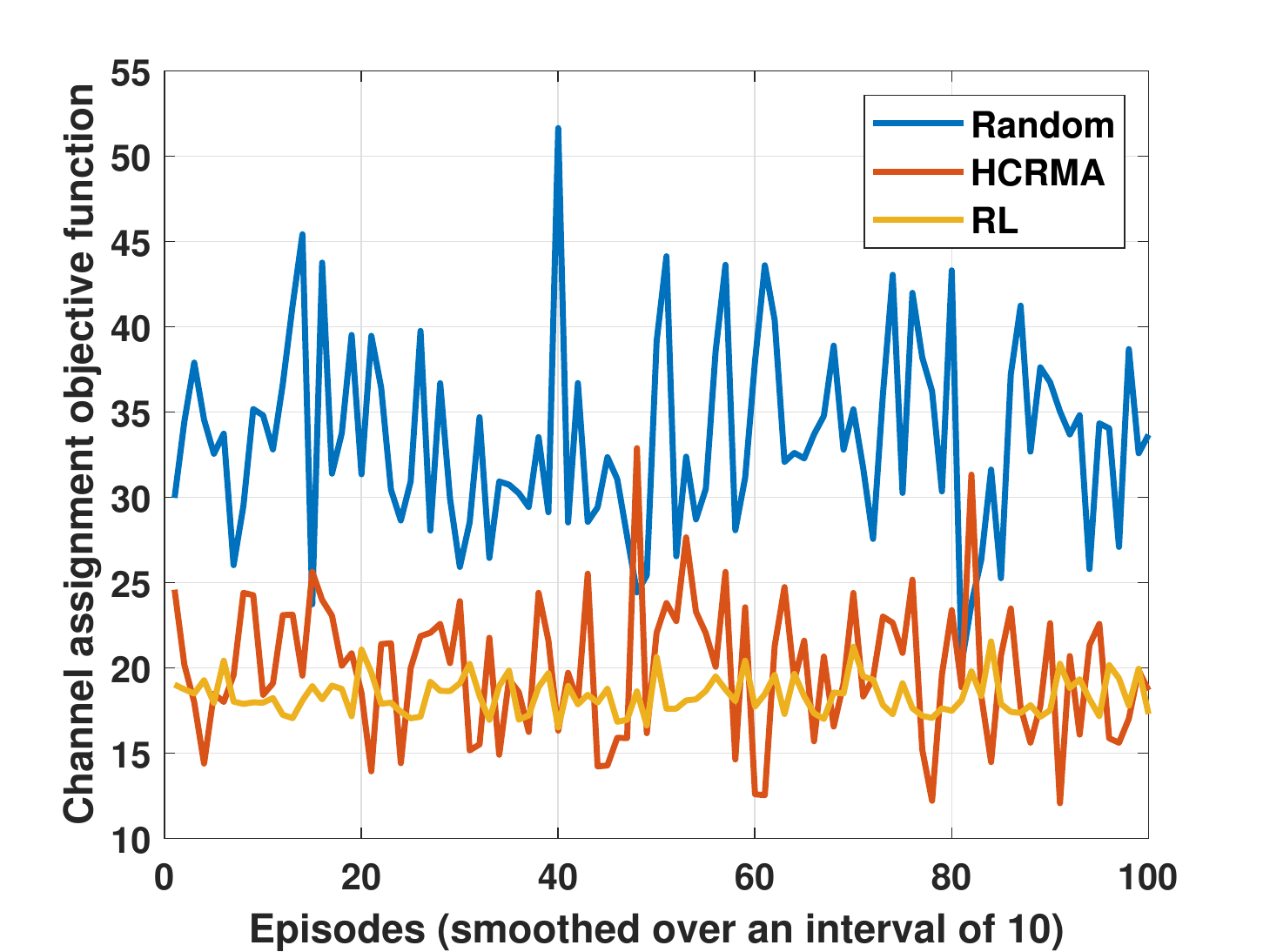}}
}
	\caption{Average penalties (channel assignment objective function) vs.  smoothed episodes ($M=2,K=6,B_{max}=10~\text{J},\Gamma=\{7,8,9\}$).}
	\label{ereq}
	\vspace{-0.5cm}
\end{figure*}
Fig.~\ref{penalties_correlated} depicts the convergence of the penalties used to learn the optimal channel assignment strategy. In fact, the RL penalties match the objective function of the optimization problem (\ref{eq:13}), which is added to the reward function as illustrated in eq. (\ref{rewardFunction}). 
\begin{figure}[H]
	\centerline{\includegraphics [width=1.05\columnwidth]  {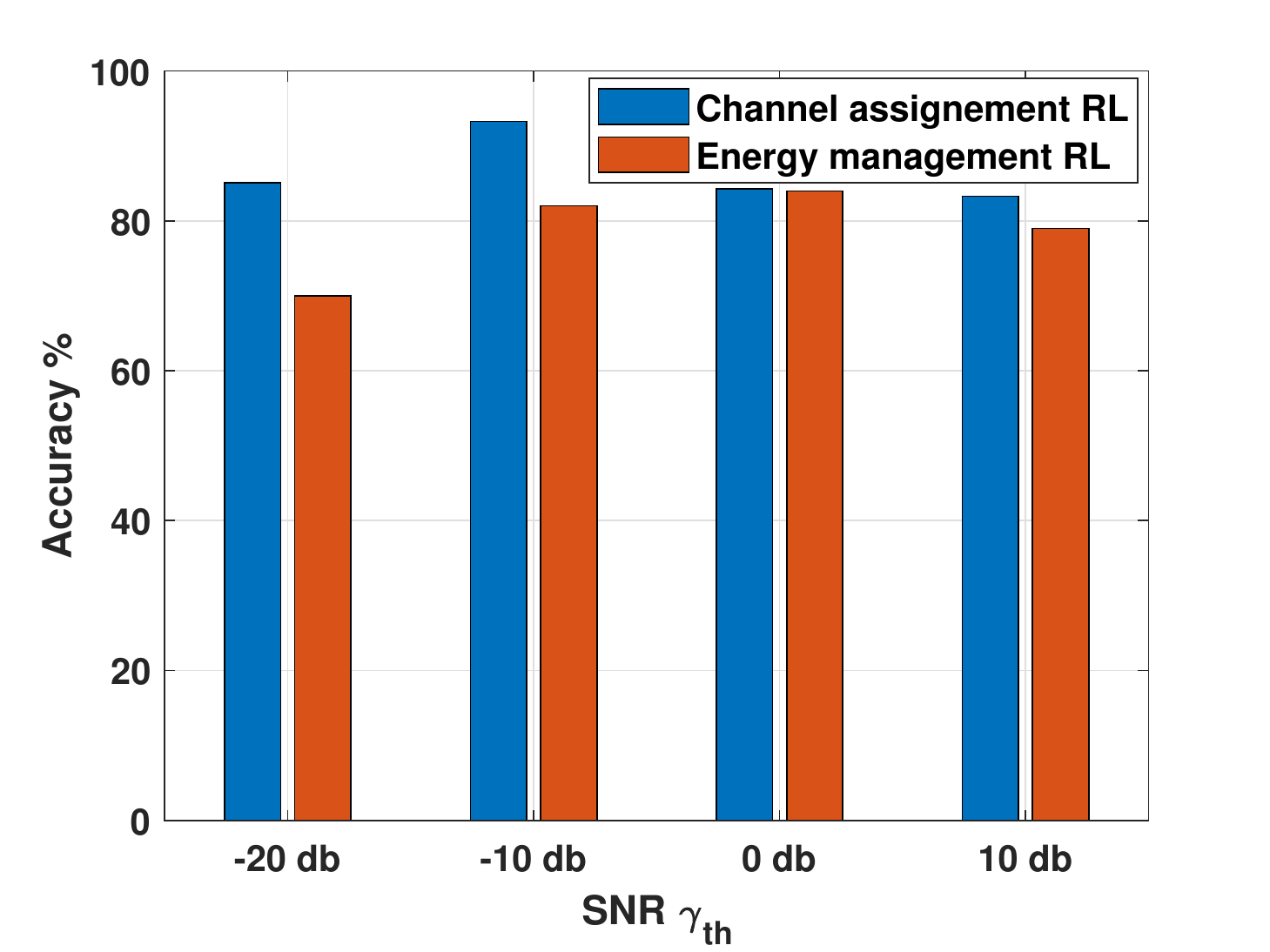}}
	\caption{Accuracy in terms of meeting constraints for both RL systems ($M=2,K=6,B_{max}=10~\text{J},\Gamma=\{7,8,9\}$).}
	\label{accuracy_2}
\end{figure}
\begin{figure}[h]
	\centerline{\includegraphics [width=1.05\columnwidth]  {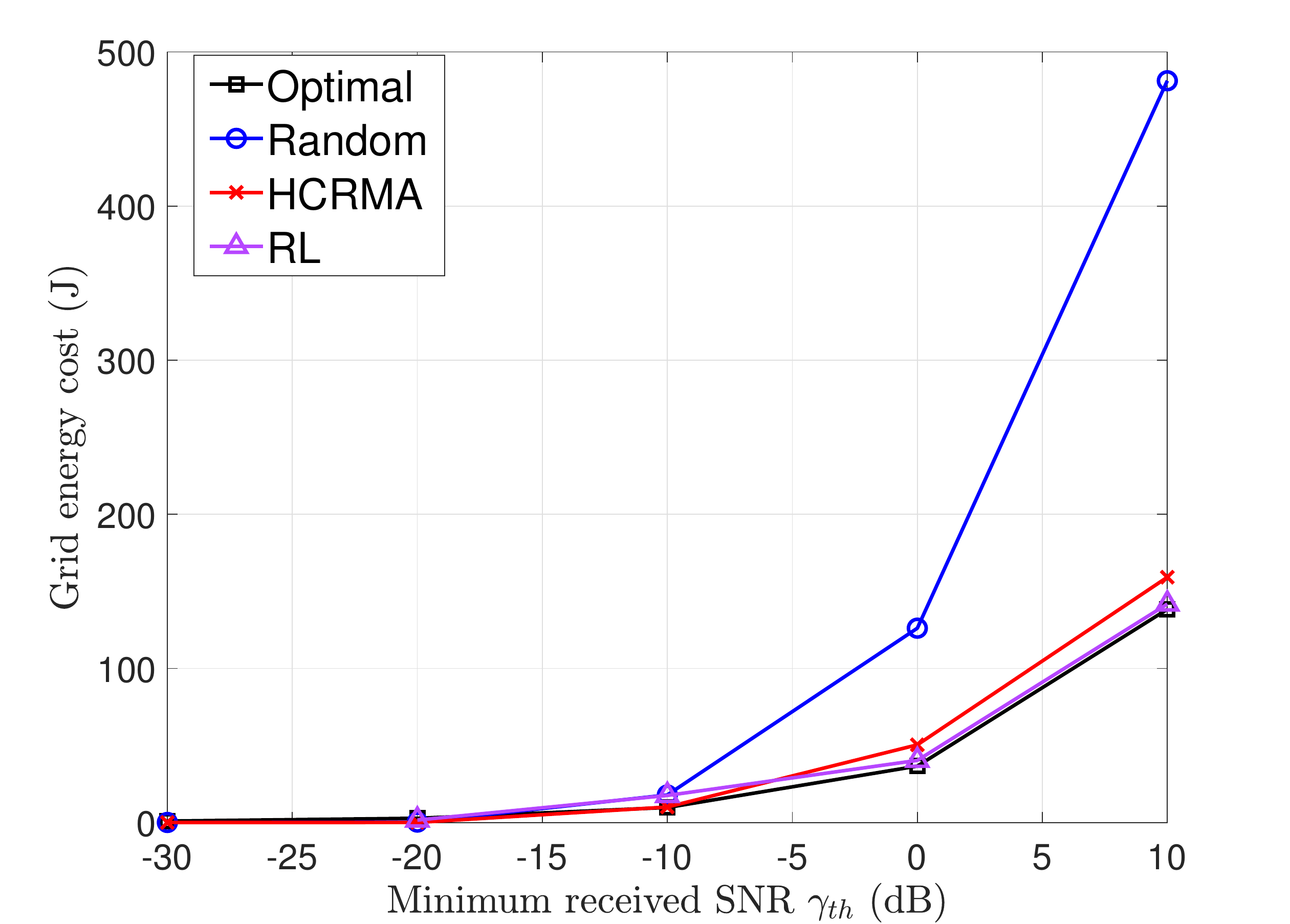}}
	\caption{Optimal grid energy cost versus SNR target ($M=2,K=6,B_{max}=10~\text{J},\Gamma=\{7,8,9\}$).}
	\label{fig4}
\end{figure}
\begin{figure}[h]
	\centerline{\includegraphics [width=1.05\columnwidth]  {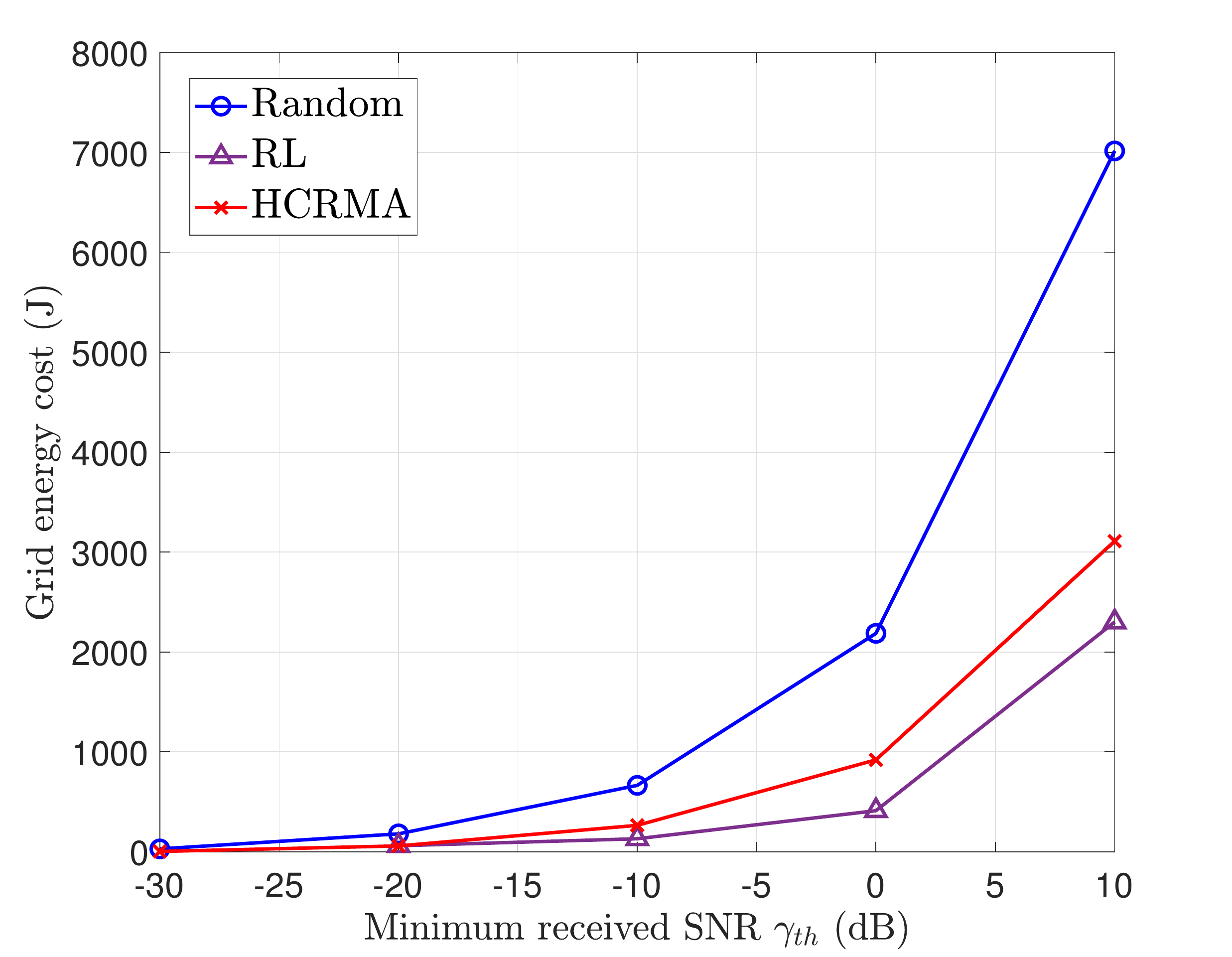}}
	\caption{Grid energy cost versus SNR target with heuristic resource management schemes ($M=3,K=20,B_{max}=200~\text{J},\Gamma=\{7,8,9,10,11,12\}$).}
	\label{fig5}
\end{figure}
\begin{figure*}[h]
\centering
	\mbox{
	     \subfigure[\label{rewards_corr}]{\includegraphics[scale=0.6]{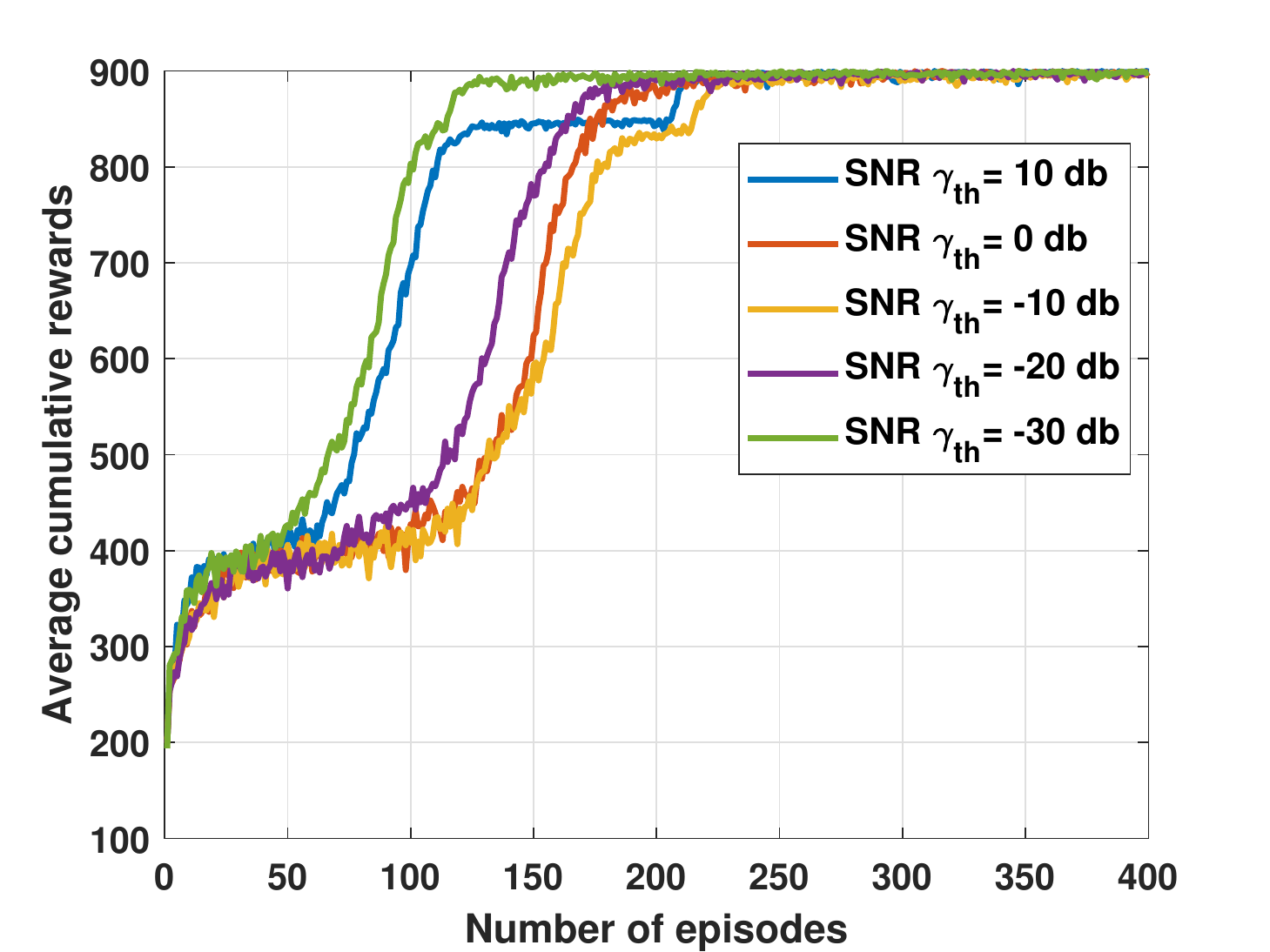}}
	     \subfigure[\label{penalties_corr}]{\includegraphics[scale=0.582]{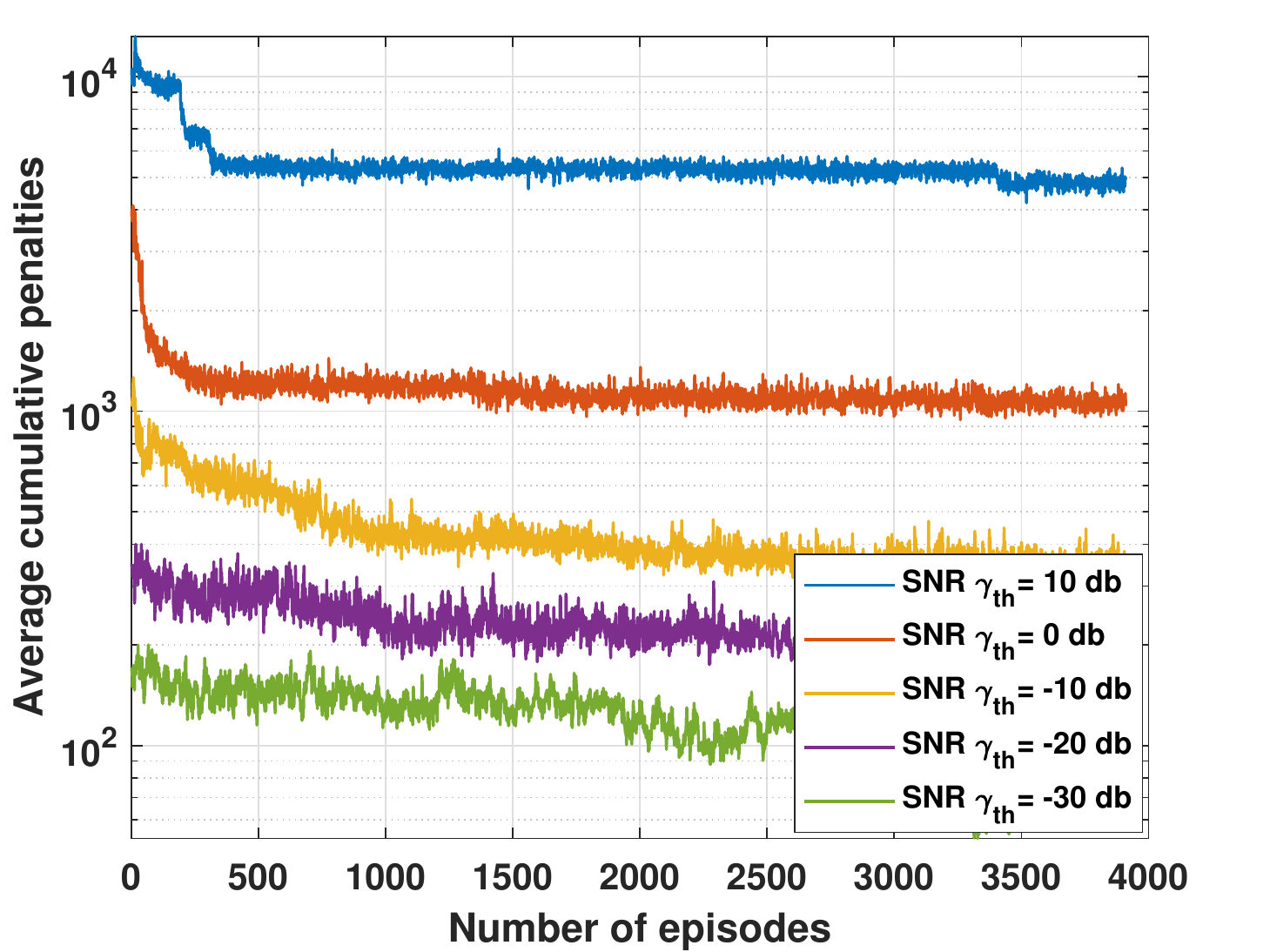}}}\\
	     \mbox{
	     \subfigure[\label{accuracy_corr}]{\includegraphics[scale=0.62]{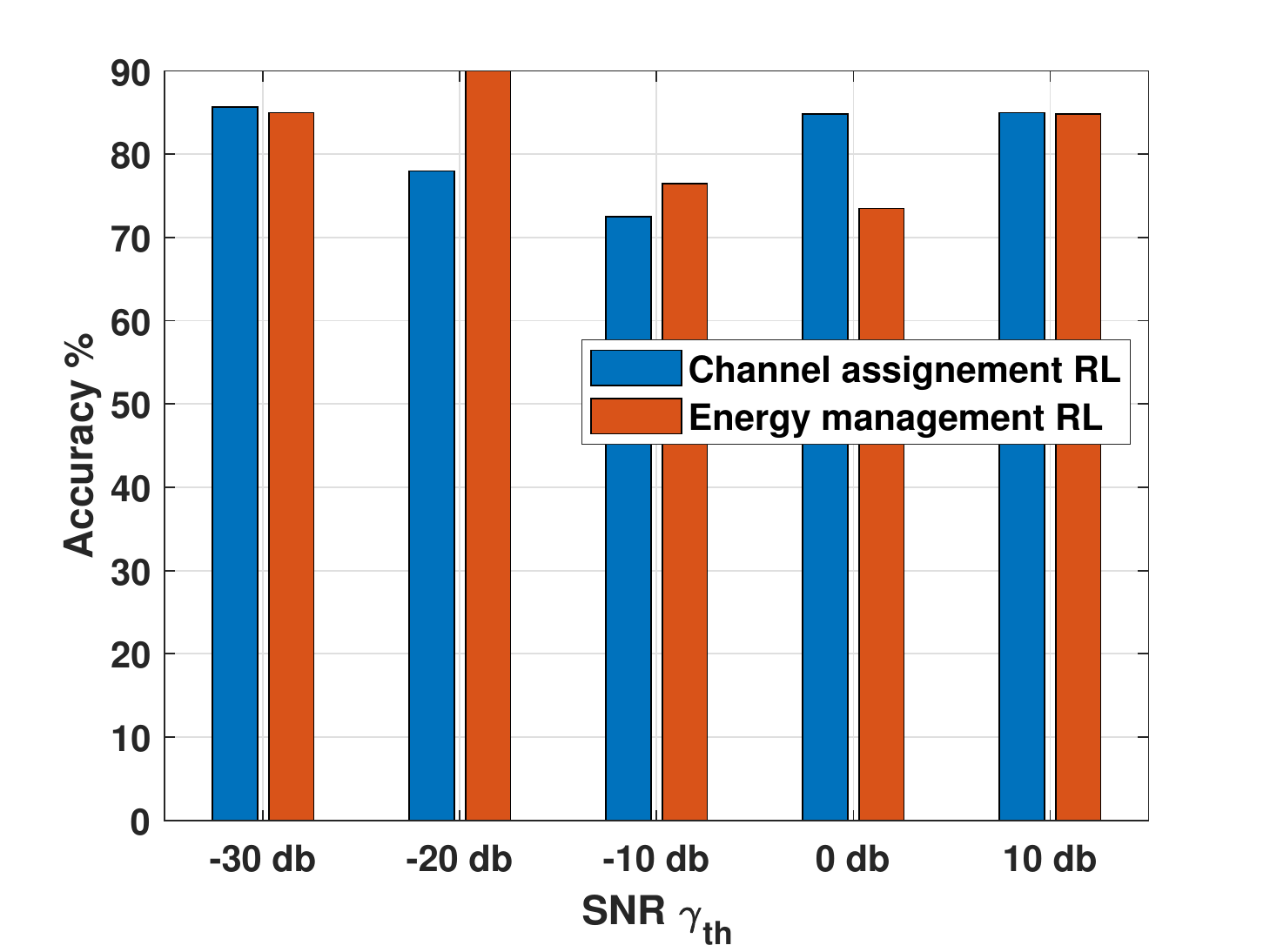}}
}
	\caption{Performance of the RL models ($M=3,K=20,B_{max}=200~\text{J},\Gamma=\{7,8,9,10,11,12\}$): (a) Average cumulative rewards vs. training episodes of the channel assignment RL system; (b) Average cumulative penalties vs. training episodes of the channel assignment RL system; (c) Accuracy in terms of meeting constraints for both RL systems.}
	\label{corr_performance}
	\vspace{-0.5cm}
\end{figure*}
According to the latter equation, the RL agent selects the optimal allocations by minimizing the penalties over the training episodes. We note that the cumulative penalties are smoothed over a window of 10. On these bases, we notice that the convergence is reached after 5000 episodes, which is not the case of the cumulative rewards that converge faster. This can be explained by the fact that the exploration space of channel allocation tasks is large. However, after learning the optimal actions related to the given states, reasonable decisions will be taken on the fly. It is worth to mention that this convergence pattern is not observed, when channels are uncorrelated due to the insignificance of the MDP process.\\
To further evaluate the performance of the channel assignment RL system over episodes, we compared it to the HCRMA heuristic and the Random resource allocation. Fig.~\ref{ereq} presents the average cumulative penalties (i.e., objective function of the channel allocation problem) over 1000 episodes for different approaches. We can see that the RL approach outperforms the heuristic in terms of allocation decisions and stability, when the SNR is equal to 10 or 0. In low SNR region, the RL presents a similar or slightly better performance.\\
\begin{figure}[H]
	\centerline{\includegraphics [width=1.05\columnwidth]  {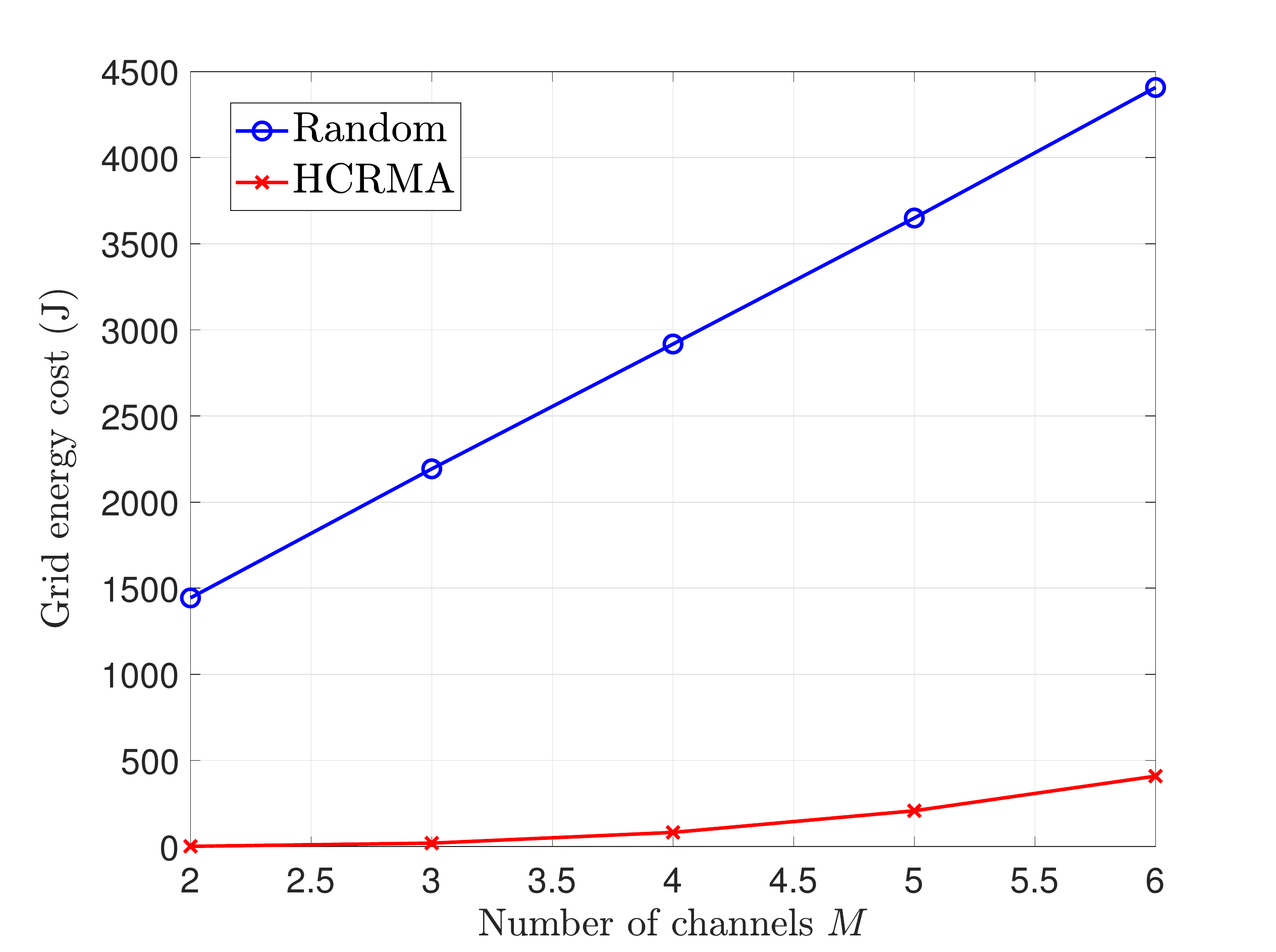}}
	\caption{Grid energy cost versus number of channels with heuristic resource management schemes ($K=40,B_{max}=200~\text{J},\Gamma=\{7,8,9,10,11,12\}, \gamma_{th}=0$ dB).}
	\label{fig6}
\end{figure}
Next, the performance of the second RL responsible for energy management is examined through evaluating its accuracy first (see Fig.~\ref{accuracy_2}), and then benchmarking it  against the low complexity algorithm HCRMA and the optimal solution  (see Fig.~\ref{fig4}). Fig.~\ref{accuracy_2} shows that both RLs have a high ability to respect different requirements and constraints of the system. Fig.~\ref{fig4} presents the grid energy cost as a function of the SNR, for different resource management approaches. Similarly to uncorrelated channel scenario, the optimal resource management solution is derived with high computational complexity. The proposed algorithm HCRMA and the RL system achieve high system performance thanks to their adequate SF and channel assignment methods. Indeed, the performance gap with the optimal is tight, for both approaches. However, in the correlated channel scenario, the RL approach outperforms the heuristic-based strategy thanks to its capacity to learn a sub-optimal allocation strategy, predict the required energy for the next frames, and manage the existing resources accordingly. This is not the case of HCRMA that executes greedy decisions.\\
We plot the performance of HCRMA and the corresponding RL system for higher number of devices $K=20$. Fig.~\ref{fig5} presents the grid energy cost versus the SNR target, for different resource management schemes. Clearly, the HCRMA  heuristic and the RL approach outperform the random scheme in terms  of grid power consumption cost for wide range of SNR. Additionally, we can see that the RL gives better results compared to the heuristic, owing to its online learning ability.  The convergence and accuracy performance of the channel assignment and energy management RL models are confirmed in Fig.~\ref{corr_performance}, where we present the average cumulative rewards and penalties, and the accuracy in terms of respecting the constraints.\\
The performance of the proposed scheme HCRMA is shown in Fig.~\ref{fig6} as a function of the number of channels $M$. The increase of the number of channels allows to us schedule more devices, which increases the total grid energy cost. HCRMA performs very well in terms of grid power consumption cost and the performance gap with the random scheme keeps increasing when the number of channels increases.\\
\begin{figure}[H]
	\centerline{\includegraphics [width=1.05\columnwidth]  {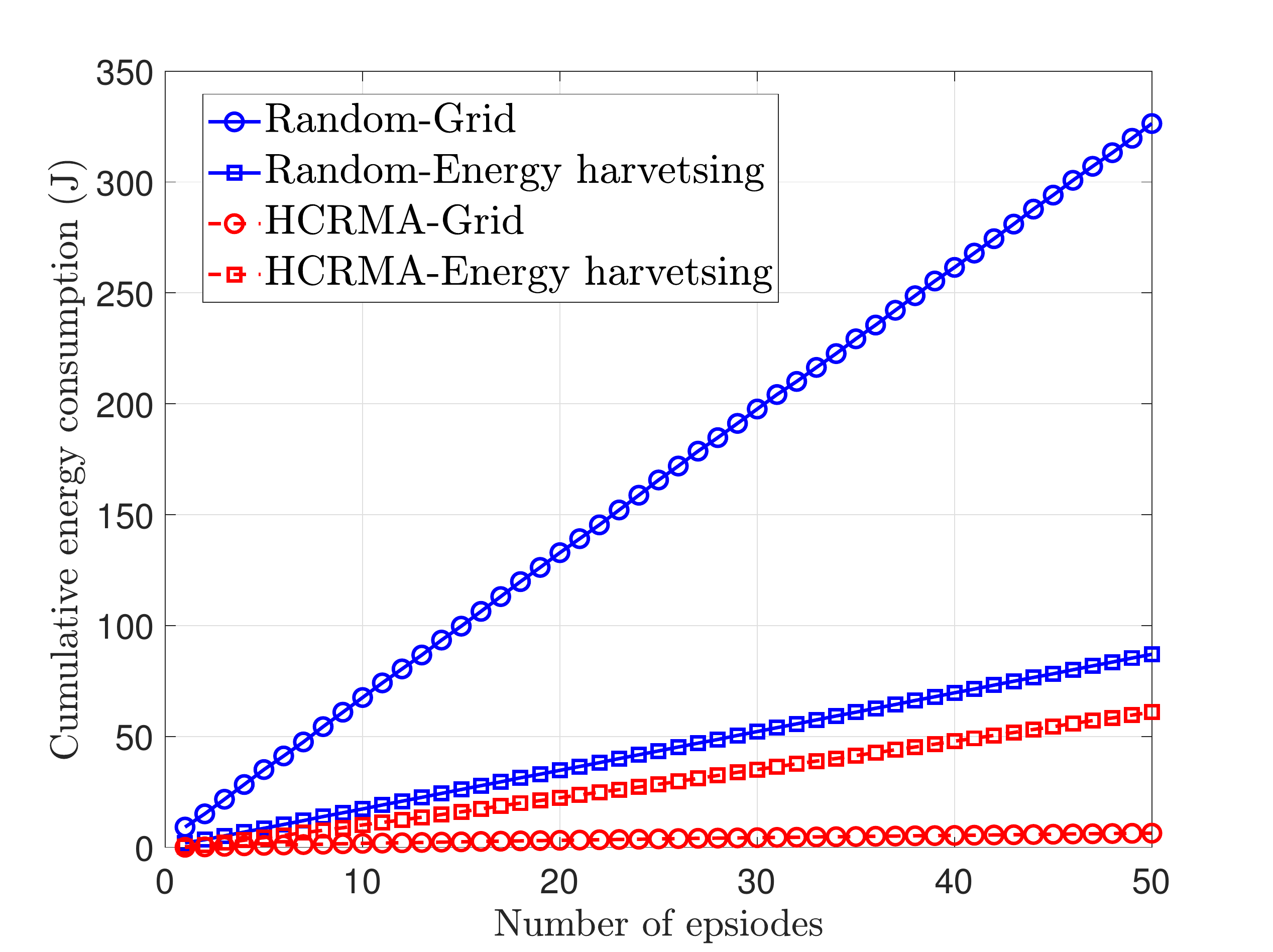}}
	\caption{Cumulative consumed grid energy and harvested energy versus number of episodes with heuristic resource management schemes. ($M=5,K=35,\gamma_{th}=-20 dB,B_{max}=200~\text{J},\Gamma=\{7,8,9,10,11,12\}$).}
	\label{fig_new2}
\end{figure}
We show in Fig.~\ref{fig_new2} the cumulative consumed grid energy and harvested energy over time. It can be seen that HCRMA allows to reduce both the consumed energy from the grid and from the battery. Moreover, HCRMA optimizes the use of the renewable energy which allows to minimize the grid energy cost.

\section{Conclusion}
This paper has investigated energy-efficient resource management in green LoRa wireless network powered by both a renewable energy source and the conventional grid. A grid power consumption minimization problem subject to the devices' quality of service demands, has been formulated. The optimal energy management solution which consists of device scheduling, SF and channel assignment, and energy management, has been solved. The online resource management has been also investigated considering both scenarios uncorrelated and time-correlated channels by developing low complexity heuristic algorithms. Moreover, smart and adaptable resource management schemes based on RL have been developed taking into account the channel and energy correlation with the objective to improve the power consumption in LoRa wireless networks. Simulation results show that the proposed resource management approaches allow efficient use of renewable energy in LoRa wireless networks.\\
Future works may cover the study of model-based RL frameworks  and their performance on LoRa systems compared to our proposed model-free RL approach. We will also focus on developing efficient resource management schemes for federated learning over LoRa wireless networks. Moreover, SWIPT system could be incorporated in LoRa wireless networks, and efficient resource management schemes may be developed.

\section*{Acknowledgment}
This work was made possible by NPRP-Standard (NPRP-S) Thirteen (13th) Cycle grant $\#$ NPRP13S-0205-200265 from the Qatar National Research Fund (QNRF) (a member of Qatar Foundation) and the TÜBITAK—QNRF Joint Funding Program grant (AICC03-0324-200005) from the Scientific and Technological Research Council of Turkey and QNRF. The findings herein reflect the work, and are solely the responsibility, of the authors.  

\balance

\end{document}